\documentclass[journal]{IEEEtran}
\usepackage{cite}
\usepackage{amsmath,amssymb,amsfonts}
\usepackage{algorithm}  
\usepackage{algpseudocode}  
\usepackage{algpseudocode}
\usepackage{graphicx}
\usepackage{epstopdf}
\usepackage{textcomp}
\usepackage{xcolor}
\usepackage{bm}
\usepackage{setspace}
\ifCLASSINFOpdf
\else
\fi
\begin{document}
%
\title{Joint Communication and Trajectory Design for Intelligent Reflecting Surface Empowered UAV SWIPT Networks}
%
%
%

\author{Zhendong~Li,~Wen~Chen,~\IEEEmembership{Senior~Member,~IEEE},~Huanqing~Cao,~Hongying~Tang,~Kunlun~Wang,~\IEEEmembership{Member,~IEEE},~and~Jun~Li,~\IEEEmembership{Senior~Member,~IEEE}
\thanks{This work is supported by National key project 2020YFB1807700 and 2018YFB1801102, by Shanghai Kewei 20JC1416502 and 22JC1404000, and NSFC 62071296. (\emph{Corresponding author: Wen Chen.})}
\thanks{Z. Li, W. Chen and H. Cao are with the Department of Electronic Engineering, Shanghai Jiao Tong University, Shanghai 200240, China (e-mail: lizhendong@sjtu.edu.cn; wenchen@sjtu.edu.cn; caohuanqing@sjtu.edu.cn).}
\thanks{H. Tang is with the Science and Technology on Microsystem Laboratory, Shanghai Institute of Microsystem and Information Technology, Chinese Academy of Sciences, Shanghai 200050, China (e-mail: tanghy@mail.sim.ac.cn).}
\thanks{K. Wang is with the School of Communication and Electronic Engineering, East China Normal University, Shanghai 200241, China. (e-mail: klwang@cee.ecnu.edu.cn).}
\thanks{J. Li is with the School of Electronic and Optical Engineering, Nanjing University of Science Technology, Nanjing 210094, China (email: jun.li@njust.edu.cn). }
}
\maketitle

\begin{abstract}
	 Aiming at the limited battery capacity of widely deployed low-power smart devices in the Internet-of-things (IoT), this paper proposes a novel intelligent reflecting surface (IRS) empowered unmanned aerial vehicle (UAV) simultaneous wireless information and power transfer (SWIPT) network framework, in which IRS is used to reconstruct the wireless channel to enhance the wireless energy transmission efficiency and coverage area of the UAV SWIPT networks. In this paper, we formulate an achievable sum-rate maximization problem by jointly optimizing UAV trajectory, successive interference cancellation (SIC) decoding order, UAV transmit power allocation, power splitting (PS) ratio and IRS reflection coefficient while taking account of user non-orthogonal multiple access (NOMA) and a non-linear energy harvesting model. Due to the coupling of optimization variables, this problem is a complex non-convex optimization problem, and it is challenging to solve it directly. We first transform the problem, and then apply the alternating optimization (AO) algorithm framework to divide the transformed problem into four sub-problems to solve it. Specifically, by applying successive convex approximation (SCA), penalty function method and difference-convex (DC) programming, UAV trajectory, SIC decoding order, UAV transmit power allocation, PS ratio and IRS reflection coefficient are alternately optimized until the convergence is achieved. Numerical simulation results verify the effectiveness of our proposed algorithm compared to other algorithms.
\end{abstract}


%
\IEEEpeerreviewmaketitle

\section{Introduction}
%
%
%
%
\IEEEPARstart{N}{owadays}, with the vigorous development of the Internet-of-things (IoT), the number of smart devices is growing rapidly \cite{8847351,8281479,8265207}. These smart devices for sending and collecting information have the characteristics of low power consumption and limited battery capacity. It is an effective way to replace the battery or charge the battery. However, if the smart devices in IoT are large-scale, such operations are time-consuming and labor-intensive, i.e., its deployment cost increases a lot.

Wireless power transfer (WPT) is a promising technology that can solve the above-mentioned challenges. The technology is flexible, easy to deploy, and does not require contact, so it has received extensive attention from industry and academia \cite{8226847,9509394,9316713}. For devices with low power consumption and limited battery capacity in IoT, wireless charging devices can dynamically join or leave the network, which is more effective. Simultaneous wireless information and power transfer (SWIPT) technology is a scheme in WPT \cite{6805330}. Through SWIPT, users can get information and energy transmission at the same time, which brings great convenience to the deployment of IoT devices. As one of the design schemes of SWIPT practical receivers, the power splitting (PS) scheme divides the signal received by the receiver into two different power streams, one part is used to decode information, and the other part is used to harvest energy. However, smart devices are usually widely distributed to collect various types of data in IoT, and traditional WPT is not efficient. Therefore, it is challenging for these smart devices to obtain energy from stationary energy stations. Although this problem can be solved by increasing the number of power stations in the target area, this will greatly increase the deployment cost of the network.

In recent years, unmanned aerial vehicles (UAVs) have been widely used in different fields. Compared with traditional fixed access points (APs), UAVs equipped with APs have the advantages of dynamic mobility, flexibility, ease of deployment and low cost \cite{8856195,8247211,8489991,8787874,7995044,7918510}. UAVs equipped with wireless energy stations can better solve the battery capacity limitation problem for widely deployed smart devices in IoT networks. Therefore, the research on UAV-enabled WPT networks has also attracted the attention of the academia \cite{8918344,9080059,8902102}. Sun \emph{et al.} investigated physical layer security enhancement methods for millimeter-wave (mmWave) UAV SWIPT networks \cite{8918344}. Wang \emph{et al.} proposed UAV-assisted non-orthogonal multiple access (NOMA) to achieve SWIPT and guarantee the secure transmission for ground passive receivers (PRs), in which the nonlinear energy harvesting model is applied \cite{9080059}. However, for UAV-assisted WPT networks, due to distance-related propagation loss, the energy transmission efficiency will decrease as the distance increases, which greatly limits the coverage of UAV-assisted SWIPT. If the wireless channel can be reconstructed and the channel gain can be increased, the coverage area of the networks can be greatly improved, which greatly stimulates the utilization of new networking paradigms to improve the performance for UAV enabled WPT networks.

Intelligent reflecting surface (IRS), also called reconfigurable intelligent surface (RIS), as a revolutionary technology, has been studied extensively by industry and academia \cite{8811733,9716123,8930608,9570775,9599533,li2021uplink}, which can reconstruct the wireless channel from the transmitter to the receiver by adjusting the amplitude and phase of the incident signal, thereby improving network performance. In detail, an IRS is an array composed of a large number of low-cost passive reflecting elements, which can be easily deployed on indoor walls or buildings. Since the IRS is a passive device, it only passively reflects the incident signal without signal processing, so it will not introduce unnecessary noise compared with the relay technology\cite{gong2019towards}. Meanwhile, compared with MIMO technology, since it is not equipped with a complex signal processing unit and it is passive, the required hardware cost and power consumption are much lower \cite{9206044}. These have greatly promoted the application of IRS in the next generation communication networks.

Based on the advantages of the IRS, the coverage and wireless energy transmission efficiency of the IRS-enabled UAV SWIPT network can be improved. Currently, the optimization and design on the IRS-assisted WPT network \cite{9650755,9531372}, the IRS-assisted UAV network \cite{9234511,9454446,9416239}, and the UAV WPT network \cite{8918344,9080059,8902102} are in progress. However, the coverage and transmission efficiency potential of these networks cannot be fully released. IRS empowered UAV SWIPT network can well address the above-mentioned challenges in the deployment of IoT devices with limited battery capacity, which is of practical significance. To the best of our knowledge, research on IRS-assisted dynamic UAV WPT networks in IoT scenarios is still in its infancy, which is novel and practical. To fully unleash the potential of drones, we consider the IRS-assisted dynamic UAV SWIPT network while considering user NOMA. In this paper, the achievable sum-rate is maximized by jointly optimizing UAV trajectory, successive interference cancellation (SIC) decoding order, UAV transmit power allocation, PS ratio and IRS reflection coefficient. Due to the high coupling of optimization variables, the concavity and convexity of the objective function and some constraints cannot be determined, so it is challenging to solve this problem directly. Therefore, we need to design an effective algorithm for IRS-assisted UAV-enabled SWIPT networks.

Based on the above background, the main contributions of this paper can be summarized as follows:
\begin{itemize}
	\item Facing the limited battery capacity of smart devices with widely deployed and low power consumption in IoT networks, we proposed an IRS empowered UAV SWIPT framework. Smart devices apply PS scheme, which allows them to harvest energy while receiving information. In addition, these devices use the NOMA scheme. Meanwhile, we formulate an achievable sum-rate maximization problem by jointly optimizing UAV trajectory, SIC decoding order, UAV transmit power allocation, PS ratio and IRS reflection coefficient. Since this problem is a complicated non-convex optimization problem, it is challenging to solve it directly.
	
	\item In order to solve the above sum-rate maximization problem, we first transform the problem, and then use the alternating optimization (AO) framework to divide the transformed problem into four sub-problems. Specifically, first, given UAV transmit power allocation, PS ratio, and IRS reflection coefficient, UAV trajectory and SIC decoding order can be jointly obtained by applying successive convex approximation (SCA) and penalty function method. Given the UAV trajectory, SIC decoding order, UAV transmit power allocation and IRS reflection coefficient, the PS ratio scheme can be obtained. Similarly, UAV transmit power allocation and IRS phase shift coefficients can also be obtained separately by using SCA, penalty function method and difference-convex (DC) programming when the other three variables are given. Finally, the four sub-problems are alternately optimized until convergence is achieved.
	
	\item Through numerical simulation, we verify the effectiveness of the proposed optimization algorithm for UAV trajectory, SIC decoding order, UAV transmit power allocation, PS ratio and IRS reflection coefficient compared with the algorithms, i.e., it can improve the achievable sum-rate of the system. For the UAV SWIPT network assisted by IRS, the sum-rate is significantly higher than that of the network without IRS assistance. Meanwhile, as the number of reflecting elements of the IRS increases, the achievable sum-rate can improve.
\end{itemize}

The remainder of this paper is organized as follows. Section II elaborates the system model and optimization problem formulation for the IRS empowered UAV SWIPT networks. Section III presents the proposed optimization algorithm for the formulated optimization problem. In Section IV, numerical results demonstrate that our algorithm has good convergence and effectiveness. Finally, conclusions are given in Section V.

\textit{Notations:} Scalars are denoted by lower-case letters, while vectors and matrices are represented by bold lower-case letters and bold upper-case letters, respectively. $\left| {x} \right|$ denotes the absolute value of a complex-valued scalar $x$, and $\left\| {\bf{x}} \right\|$ denotes the Euclidean norm of a complex-valued vector $\bf{x}$. $diag(\bf{x})$ denotes a diagonal matrix whose diagonal elements are the corresponding elements in vector $\bf{x}$. For a square matrix $\bf{X}$, $\rm{tr(\bf{X})}$, $\rm{rank(\bf{X})}$, ${\bf{X}}^H$ and ${\bf{X}}_{m,n}$ denote its trace, rank, conjugate transpose and $m,n$-th entry, respectively, while ${\bf{X}} \succeq 0$ represents that $\bf{X}$ is a positive semidefinite matrix. ${\mathbb{C}^{M \times N}}$ denotes the space of ${M \times N}$ complex matrices. $j$ denotes the imaginary unit, i.e., $j^2=-1$. $\mathbb{E}\left\{  \cdot  \right\}$ represents the expectation of random variables. Finally, the distribution of a circularly symmetric complex Gaussian (CSCG) random vector with mean $\mu$ and covariance matrix $\bf{C}$ is denoted by $ {\cal C}{\cal N}\left( {\mu,\bf{C}} \right)$, and $\sim$ stands for `distributed as'.
\section{System Model and Problem Formulation}
\subsection{System Model}
In this paper, we consider an IRS empowered downlink UAV SWIPT network in IoT consisting of a rotary-wing UAV with a single omni-directional antenna, an IRS and $K$ single antenna users (i.e., smart devices with limited battery capacity). As shown as in Fig. 1, the UAV simultaneously sends signals to users. The IRS can be deployed on a building to assist the SWIPT networks from the UAV to users, which is equipped with a uniform linear array (ULA) of $M$ reflecting elements\footnote{It is worth noting that the algorithm proposed in this paper can be extended to the IRS equipped with a uniform planar array (UPA) by considering the corresponding antenna array response.}. Meanwhile, the IRS is also equipped with a smart controller, which coordinates the UAV and IRS for both channel acquisition and information and power transmission. We assume that all channels in this paper are quasi-static flat-fading and the channel state information (CSI) of all channels is perfectly known at the UAV. For the channel estimation of IRS empowered UAV networks, we can use state-of-the-art algorithms such as direct cascaded channel estimation \cite{8879620} and separable cascaded channel estimation \cite{9361077}.
\begin{figure}
	\centerline{\includegraphics[width=7.0cm]{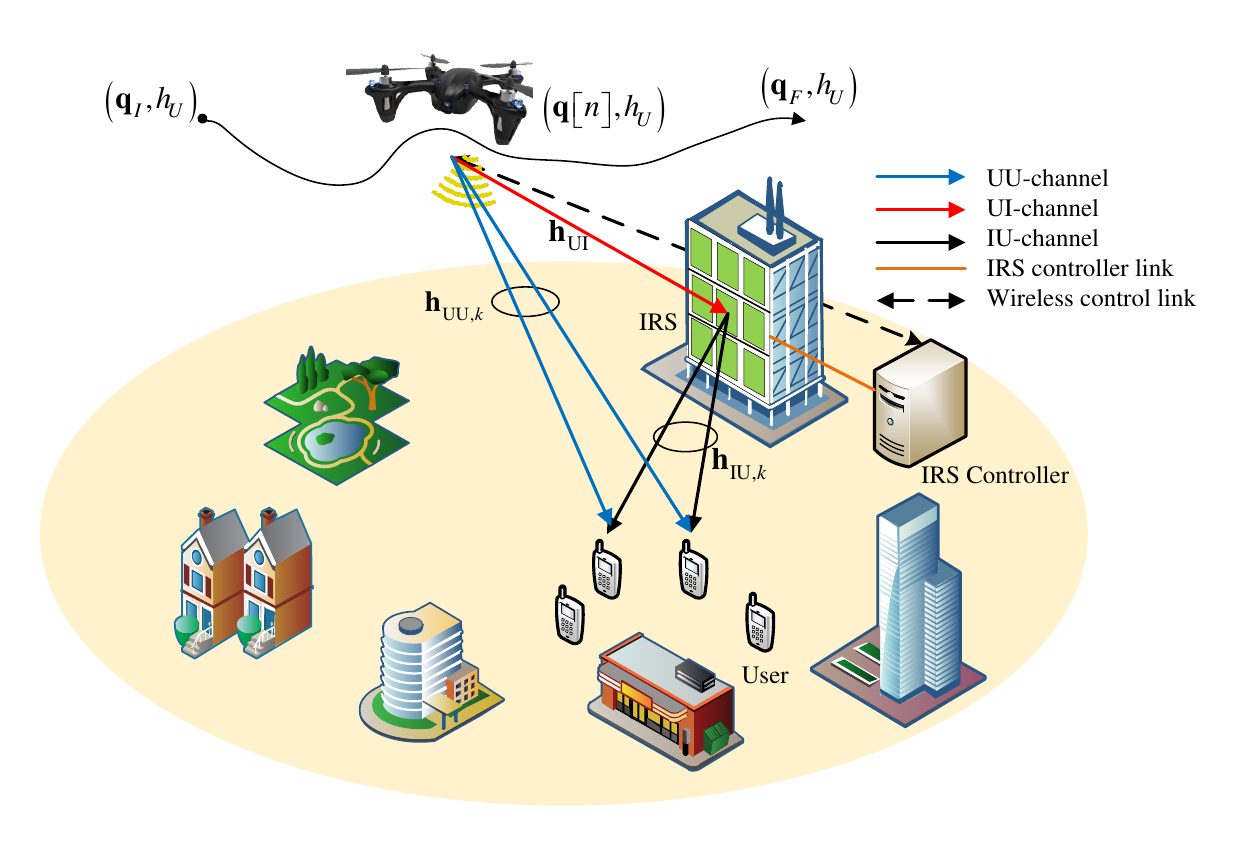}}
	\caption{IRS empowered UAV SWIPT networks.}
	\label{Fig1}
\end{figure}

Without loss of generality, we consider a 3D Cartesian coordinate system, where the $k$-th single-antenna user's coordinate is ${{\bf{w}}_k} = {\left[ {{x_k},{y_k}},0 \right]^T}$. The UAV flies at a fixed altitude ${h_u}$, which is the minimum altitude to avoid any collision with building. We consider a finite time period $T$ to guarantee the efficiency of simultaneous information and power transmission. For simplicity, the time period $T$ is divided into $N$ time slots, indexed by $n=1,...,N$. Each time slot $\delta  = \frac{T}{N}$ is selected to be small enough to ensure that the UAV position is approximately unchanged when flying at the maximum speed ${V_{\max }}$. Hence, the 3D trajectory of UAV can be approximated by ${\bf{q}}\left[ n \right] = {\left[ {x\left[ n \right],y\left[ n \right]},h_u \right]^T},n=1,...,N$. We consider the initial and final position of the UAV can be denoted by ${{\bf{q}}_I}$ and ${{\bf{q}}_F}$, respectively. The trajectory of UAV should satisfy the following constraints
\begin{equation}
{{\bf{q}}_I} = {\bf{q}}\left[ 1 \right],
\end{equation}
\begin{equation}
{{\bf{q}}_F} = {\bf{q}}\left[ N \right],
\end{equation}
\begin{equation}
{\left\| {{\bf{q}}\left[ {n + 1} \right] - {\bf{q}}\left[ n \right]} \right\|^2} \le {\left( {{V_{\max }}\delta } \right)^2},n=1,...,N-1.
\end{equation}

For the IRS with $M$ reflecting elements, each element reflects the received signal from the UAV with an adjustable amplitude and phase shift. Let $\bm{\theta} \left[ n \right] = {\left[ {{\theta _1}\left[ n \right],...,{\theta _M}\left[ n \right]} \right]^T} \in {\mathbb{C}^{M \times 1}},\forall n$, and we model the IRS reflection matrix by using ${\bf{\Theta }}\left[ n \right] = diag\left( {{\beta _1}\left[ n \right]{e^{j{\theta _1}\left[ n \right]}},...,{\beta _M}\left[ n \right]{e^{j{\theta _M}\left[ n \right]}}} \right) \in \mathbb{C} {^{M \times M}}$, where ${\theta _m}\left[ n \right] \in \left[ {0,2\pi } \right),\forall m,n$ and $\beta _m \left[ n \right] \in \left[ {0,1} \right], \forall m,n$ denote the phase shift and amplitude reflection coefficient of the $m$-th IRS element, respectively. For simplicity, we set ${\beta _m\left[ n \right] } = 1$ to achieve the maximum reflecting power gain. In the $n$-th time slot, the phase shift of the $m$-th element should satisfy the following constraint
\begin{equation}
0 \le {\theta _m}\left[ n \right] < 2\pi ,\forall m,n.
\end{equation}
Furthermore, the first element of the IRS is regarded as the reference point whose 3D coordinate can be denoted by ${\bf{w}}_r = {\left[ x_r,y_r,h_r\right]^T}$, respectively. Hence, the distance between the IRS and the UAV or the ground users can be approximated by that between the reference point and the corresponding node. The channel gain from the UAV to the IRS, from the UAV to the $k$-th ground user, and from the IRS to the $k$-th ground user can be denoted by ${{\bf{h}}_{{\rm{UI}}}} \in \mathbb{C} {^{M \times 1}}$, ${{h}_{{\rm{UU}},k}} \in \mathbb{C}$, and ${{\bf{h}}_{{\rm{IU}},k}} \in \mathbb{C} {^{M \times 1}}$, respectively. 

Since the UAV usually flies at a high altitude and the IRS is commonly placed on the building, the channel from the UAV to the IRS (i.e., UI-channel) can be modeled to be a line-of-sight (LoS) channel. Since the UAV is flying or hovering at a certain height in the air, the channel from the UAV to the $k$-th user (i.e., UU-channel) has both LoS and non-line-of-sight (NLoS) components, so we model it as a Rician channel \cite{9095281}. Besides, the channel from the IRS to the $k$-th ground user (i.e., IU-channel) can also be modeled by a Rician fading channel. The LoS component of the channel associated with the IRS can be expressed by the responses of the ULA. The array response of $M$-element ULA of the IRS in the $n$-th time slot can be given by
\begin{equation}
{\bf{h}}\left[ n \right] = {\left[ {1,{e^{ - j2\pi \frac{d}{\lambda }\cos \phi \left[ n \right]}},...,{e^{ - j2\pi \frac{d}{\lambda }\left( {M - 1} \right)\cos \phi \left[ n \right]}}} \right]^T},\forall n,
\end{equation}
where $\phi\left[ n \right] $ is angle-of-arrival (AoA) from the UAV to the IRS in the $n$-th time slot, $\cos \phi \left[ n \right] = \frac{{{x_r} - x\left[ n \right]}}{{{d_{{\rm{UI}}}}\left[ n \right]}}$ with ${d_{{\rm{UI}}}}\left[ n \right] = \left\| {{\bf{q}}\left[ n \right] - {{\bf{w}}_r}} \right\|$ denotes the distance from the UAV to the IRS in the $n$-th time slot, $\lambda $ is the carrier wavelength, and $d$ represents the array interval. Therefore, the channel gain of the UI-channel in the $n$-th time slot can be denoted by 
\begin{equation}
{{\bf{h}}_{{\rm{UI}}}}\left[ n \right] = \sqrt {\frac{{{\beta _0}}}{{{{\left( {{d_{{\rm{UI}}}}\left[ n \right]} \right)}^2}}}} {\rm{ }}{\bf{h}}\left[ n \right],\forall n,
\end{equation}
where $\beta_0 $ is the path loss when the reference distance is 1m. In addition, the channel gain of the UU-channel in the $n$-th time slot can be expressed by
\begin{equation}
	\begin{aligned}
		{h_{{\rm{UU}},k}}\left[ n \right] = \sqrt {\frac{{{\beta _0}}}{{{{\left( {{d_{{\rm{UU}},k}}\left[ n \right]} \right)}^\alpha }}}} \left({ \sqrt {\frac{{{\kappa _1}}}{{1 + {\kappa _1}}}} h_{{\rm{UU}},k}^{{\rm{LoS}}}\left[ n \right] }\right.\\
		\left.{+ \sqrt {\frac{1}{{1 + {\kappa _1}}}} h_{{\rm{UU}},k}^{{\rm{NLoS}}}\left[ n \right]} \right),\forall k,n,
	\end{aligned}
\end{equation} 
where $\alpha $ is the corresponding path loss exponent related to the UU-channel and $\kappa_1 $ is the Rician factor\footnote{In fact, the Rician factor is related to the trajectory of the UAV \cite{8698468}. Since the channel gain between the UAV and the user is largely dependent on the path loss caused by the distance, for the convenience of analysis, we assume that the Rician factor in each time slot is fixed.}. ${d_{{\rm{UU}},k}}\left[ n \right] = \left\| {{\bf{q}}\left[ n \right] - {{\bf{w}}_k}} \right\|$ denotes the distance from the UAV to the $k$-th user in the $n$-th time slot. $h_{{\rm{UU}},k}^{{\rm{LoS}}}\left[ n \right] = 1$ and $h_{{\rm{UU}},k}^{{\rm{NLoS}}}\left[ n \right] \sim {\cal C}{\cal N}\left( {0,1} \right)$ represents the random scattering component. Similarly, the channel gain of the IU-channel can be denoted by 
\begin{equation}
	\begin{aligned}
		{{\bf{h}}_{{\rm{IU}},k}} = \sqrt {\frac{{{\beta _0}}}{{{{\left( {{d_{{\rm{IU}},k}}} \right)}^\gamma }}}} \left( {\sqrt {\frac{{{\kappa _2}}}{{1 + {\kappa _2}}}} {\rm{ }}{\bf{h}}_{{\rm{IU}},k}^{{\rm{LoS}}} + \sqrt {\frac{1}{{1 + {\kappa _2}}}} {\bf{h}}_{{\rm{IU}},k}^{{\rm{NLoS}}}} \right),\forall k,
	\end{aligned}
\end{equation}
where $\gamma $ is the corresponding path loss exponent related to the IU-channel, and $\kappa_2 $ is the Rician factor. ${d_{{\rm{IU}},k}} = \left\| {{{\bf{w}}_r} - {{\bf{w}}_k}} \right\|$ denotes the distance from the IRS to the $k$-th user in the $n$-th time slot. The LoS component ${\bf{h}}_{{\rm{IU}},k}^{{\rm{LoS}}} \in \mathbb{C} {^{M \times 1}}$ can be denoted by 
\begin{equation}
{\bf{h}}_{{\rm{IU}},k}^{{\rm{LoS}}} = \left[ {1,{e^{ j2\pi \frac{d}{\lambda }\cos\varphi_k}},...,{e^{ j2\pi \frac{d}{\lambda }\left( {M - 1} \right)\cos\varphi_k}}} \right]^T,\forall k,
\end{equation}
where $\varphi_k$ is the angle-of-departure (AoD) from the IRS to the $k$-th user, $\cos {\varphi _k} = \frac{{{x_k} - {x_r}}}{{{d_{{\rm{IU}},k}}}}$. The NLoS component can be denoted by ${\bf{h}}_{{\rm{IU}},k}^{{\rm{NLoS}}} \sim {\cal C}{\cal N}\left( {{\bf{0}},{{\bf{I}}_M}} \right)$. Therefore, with the aid of the IRS, the combined channel power gain from the UAV to the $k$-th user in the $n$-th time slot can be expressed as
\begin{equation}
	{H_k}\left[ n \right] = {\left| {{h_{{\rm{UU}},k}}\left[ n \right] + {\bf{h}}_{{\rm{IU}},k}^H{\bf{\Theta }}\left[ n \right]{{\bf{h}}_{{\rm{UI}}}}\left[ n \right]} \right|^2},\forall k,n.
\end{equation}

In this paper, we assume that all uses share the same frequency, thus the UAV applies NOMA to provide communication for the users. specifically, the transmission signal of UAV by invoking superposition coding (SC) can be denoted as $\tilde s = \sum\limits_{k = 1}^K {\sqrt {{p_k}\left[ n \right]} } {s_k}\left[ n \right]$, where ${s_k}\left[ n \right] \sim {\cal C}{\cal N}\left( {0,1} \right)$ denotes the transmission data sent by UAV to the $k$-th user, and ${{p_k}\left[ n \right]}$ denotes the transmit power allocated to the $k$-th user in the $n$-th time slot. Without loss of generality, it should satisfy the following constraints
\begin{equation}
	{p_k}\left[ n \right] \ge 0,\forall k,n,
\end{equation}
\begin{equation}
	\sum\limits_{k = 1}^K {{p_k}\left[ n \right]}  \le {P_{\max }},\forall n,
\end{equation}
where ${P_{\max }}$ denotes the maximum transmit power of the UAV.

Therefore, the signal received by the $k$-th user in the $n$-th slot can be expressed as
\begin{equation}
	\begin{aligned}
		&{y_k}\left[ n \right] = \underbrace {\left( {{h_{{\rm{UU}},k}}\left[ n \right] + {\bf{h}}_{{\rm{IU}},k}^H{\bf{\Theta }}\left[ n \right]{{\bf{h}}_{{\rm{UI}}}}\left[ n \right]} \right) {\sqrt {{p_k}\left[ n \right]} } {s_k}\left[ n \right]}_{{\rm{desired~signal}}} +\\
		& \underbrace {\left( {{h_{{\rm{UU}},k}}\left[ n \right] + {\bf{h}}_{{\rm{IU}},k}^H{\bf{\Theta }}\left[ n \right]{{\bf{h}}_{{\rm{UI}}}}\left[ n \right]} \right)\sum\limits_{i \ne k}^K {\sqrt {{p_i}\left[ n \right]} } {s_i}\left[ n \right]}_{{\mathop{\rm int}} {\rm{erference~signal}}} + {n_k},\forall k,n,
	\end{aligned}
\end{equation}
where ${n_k} \sim {\cal C}{\cal N}\left( {0,\sigma _k^2} \right)$ is additive white Gaussian noise (AWGN).

According to the NOMA mechanism, each user adopts SIC to remove inter-user interference. Specifically, a user with a stronger channel power gain first decodes the signal of a user with a weaker power channel gain before decoding its own signal. Therefore, the decoding order needs to be paid more attention in NOMA. Herein, we introduce a set of binary variables ${\psi _{i,k}}\left[ n \right] \in \left\{ {0,1} \right\},\forall n,i \ne k$ to describe the decoding order among users. If the channel power gain of the $i$-th user is stronger than that of the $k$-th user in the $n$-th time slot, we let ${\psi _{i,k}}\left[ n \right] = 1$. Otherwise, ${\psi _{i,k}}\left[ n \right] = 0$. Therefore, it should satisfy the following constraints:
\begin{equation}
	{\psi _{i,k}}\left[ n \right] = \left\{ {\begin{array}{*{20}{c}}
			{1,{\rm{ ~}}if{\rm{~}}{H_i}\left[ n \right] > {H_k}\left[ n \right]}\\
			{0,{\rm{~~~~~~~}}otherwise{\rm{  ~~~  }}}
	\end{array}} \right.,
\end{equation}
\begin{equation}
	{\psi _{i,k}}\left[ n \right] + {\psi _{k,i}}\left[ n \right] = 1,\forall n,i \ne k.
\end{equation}
Moreover, for a given decoding order, the UAV transmit power allocation should satisfy
\begin{equation}
	{p_k}\left[ n \right] \ge {\psi _{i.k}}{p_i}\left[ n \right],\forall n,i \ne k,
\end{equation}
which ensures that higher power needs to be allocated to users with weaker channel power gain \cite{8114722}. This communication method can well guarantee the fairness of users, i.e., it can well guarantee the communication of users with weaker channel conditions.
\begin{figure}
	\centerline{\includegraphics[width=9cm]{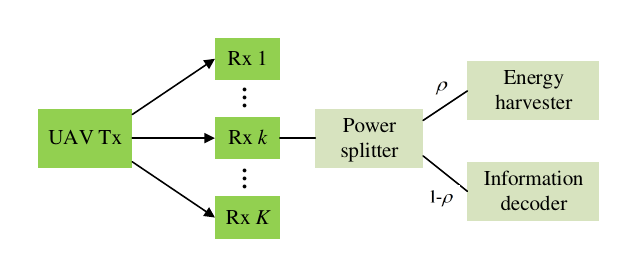}}
	\caption{PS receiver architecture.}
	\label{Fig2}
\end{figure}

In addition, we consider the power splitting (PS) receiver architecture at the users for information decoding and energy harvesting, which is actually one of the most widely used architectures in SWIPT networks. The architecture diagram is shown in the Fig. 2. Specifically, In the $n$-th time slot, the radio frequency (RF) signal received by the $k$-th ground user is split with a PS ratio ${\rho _k}\left[ n \right]$, which should satisfy the following constraint
\begin{equation}
0 \le {\rho _k}\left[ n \right] \le 1,\forall k,n,
\end{equation}
which represents the PS ratio of each ground user in the $n$-th time slot should be between zero and one. The ${\rho _k\left[ n \right]}$ portion of the received power is used by the $k$-th user for energy harvesting in the $n$-th time slot, and the remaining $(1-{\rho _k\left[ n \right]})$ portion is used by the $k$-th user for decoding information in the $n$-th time slot. Hence, the signal split to the $k$-th user in the $n$-th time slot for decoding information can be denoted by
\begin{equation}
	\begin{aligned}
		y_k^{{\rm{ID}}}\left[ n \right] = \sqrt {1 - {\rho _k}\left[ n \right]} \left( {\left( {{h_{{\rm{UU}},k}}\left[ n \right] + {\bf{h}}_{{\rm{IU}},k}^H{\bf{\Theta }}\left[ n \right]{{\bf{h}}_{{\rm{UI}}}}\left[ n \right]} \right)}\right.\\
		\left.{\sum\limits_{i = 1}^K  \sqrt {{p_i}\left[ n \right]} {s_i}\left[ n \right] + {n_k}} \right) + {z_k},\forall k,n,
	\end{aligned}
\end{equation}
where ${z_k} \sim {\cal C}{\cal N}\left( {0,\delta _k^2} \right)$ denotes the noise introduced by information decoder for the $k$-th user.

Therefore, the received signal to interference plus noise ratio (SINR) of the $k$-th user by applying SIC for information decoding in the $n$-th slot can be expressed as
\begin{equation}
	\begin{aligned}
		{\rm{SIN}}{{\rm{R}}_k}\left[ n \right]\! =\! \frac{{\left( {1 \!-\! {\rho _k}\left[ n \right]} \right){p_k}\left[ n \right]{H_k}\left[ n \right]}}{{\left( {1\! - \!{\rho _k}\left[ n \right]} \right)\!\left( {\sum\limits_{i \ne k}^K {{\psi _{i,k}}\left[ n \right]{p_i}\left[ n \right]{H_k}\left[ n \right]}\!  + \!\sigma _k^2} \right)\! \!+ \!\delta _k^2}},\\
		\forall n,i \ne k,
	\end{aligned}	
\end{equation}Therefore, the achievable rate (bps/Hz) of the $k$-th user in the $n$-th time slot can be expressed as
\begin{equation}
	{R_k}\left[ n \right] = {\log _2}\left( {1 + {\rm{SIN}}{{\rm{R}}_k}\left[ n \right]} \right),\forall k,n.
\end{equation}
The achievable sum-rate in the time period $T$ can be expressed as
\begin{equation}
{R_{{\rm{sum}}}} = \sum\limits_{n = 1}^N {\sum\limits_{k = 1}^K {{R_k}\left[ n \right]} } .
\end{equation}
In addition, the signal split to the $k$-th user in the $n$-th time slot for harvesting energy can be denoted by
\begin{equation}
	\begin{aligned}
		y_k^{{\rm{EH}}}\left[ n \right] = \sqrt {{\rho _k}\left[ n \right]} \left( {\left( {{h_{{\rm{UU}},k}}\left[ n \right] + {\bf{h}}_{{\rm{IU}},k}^H{\bf{\Theta }}\left[ n \right]{{\bf{h}}_{{\rm{UI}}}}\left[ n \right]} \right)}\right.\\
		\left.{\sum\limits_{i = 1}^K  \sqrt {{p_i}\left[ n \right]} {s_i}\left[ n \right] + {n_k} }\right),\forall k,n.
	\end{aligned}
\end{equation}
Hence, the received power of the $k$-th user in the $n$-th time slot can be given by
\begin{equation}
	{P_k}\left[ n \right] = {\rho _k}\left[ n \right]\left( {\sum\limits_{i = 1}^K {{p_i}\left[ n \right]{H_k}\left[ n \right]}  + \sigma _k^2} \right),\forall k,n.
\end{equation}
Furthermore, in order to accurately describe the energy harvesting, this paper adopts the non-linear energy harvesting model based on the practical system, thus the harvesting power of the $k$-th user in the $n$-th time can be given by
\begin{equation}
	\Xi \left( {{P_k}}\left[ n \right] \right)\! =\! \left( {\frac{{{\xi _k}}}{{{X_k}\left( {1 \!+ \!\exp \left( { - {a_k}\left( {{P_k\left[ n \right]} \!-\! {b_k}} \right)} \right)} \right)}} \!-\! {Y_k}} \right),\!\forall k,n,
\end{equation}
where $\xi _k$ denotes the maximum power that the $k$-th user can harvest, ${a_k}$ and ${b_k}$ are parameters related to specific circuit specifications, ${X_k} = {{\exp \left( {{a_k}{b_k}} \right)} \mathord{\left/
		{\vphantom {{\exp \left( {{a_k}{b_k}} \right)} {\left( {1 + \exp \left( {{a_k}{b_k}} \right)} \right)}}} \right.
		\kern-\nulldelimiterspace} {\left( {1 + \exp \left( {{a_k}{b_k}} \right)} \right)}}$ and ${Y_k} = {{{\xi _k}} \mathord{\left/
		{\vphantom {{{\xi _k}} {\exp \left( {{a_k}{b_k}} \right)}}} \right.
		\kern-\nulldelimiterspace} {\exp \left( {{a_k}{b_k}} \right)}}$ \cite{7934322}. Then the energy harvesting of the $k$-th user in the $n$-th time slot can be expressed as
\begin{equation}
	{E_k\left[ n \right]} = \delta \Xi \left( {{P_k\left[ n \right]}} \right),\forall k, n,
\end{equation}
In order to meet the energy constraint of the $k$-th user in the $n$-th time slot, ${E_k\left[ n \right]}$ needs to meet the following constraint
\begin{equation}
{E_k\left[ n \right]} \ge \chi_{th} ,\forall k,n,
\end{equation}
where $\chi_{th}$ is the energy threshold of the $k$-th user in the $n$-th time slot. It can be seen from Eq. (20) and (26) that there is a tradeoff between the achievable rate and energy harvesting for the $k$-th user in the $n$-th time slot. 

\subsection{Problem Formulation}
In this paper, we maximize the achievable sum-rate for all users in the IRS empowered UAV SWIPT networks by jointly optimizing UAV trajectory ${\bf{Q}} = \left\{ {{\bf{q}}\left[ n \right],\forall n} \right\}$, SIC decoding order ${\bm{\psi }} = \left\{ {{\psi _{i,k}}\left[ n \right],\forall n,i \ne k} \right\}$, UAV transmit power allocation ${\bf{p}} = \left\{ {p_k\left[ n \right],\forall k, n} \right\}$, PS ratio ${\bm{\rho }} = \left\{ {{\rho _k}\left[ n \right],\forall k,n} \right\}$ and IRS reflection coefficient ${\bm{\theta }} = \left\{ {{\theta _m}\left[ n \right],\forall m,n} \right\}$. The optimization problem can be formulated as follows
\begin{subequations}
	\begin{align}
		{\cal P}0:~~~~&\mathop {{\rm{max}}}\limits_{{\bf{Q}},{\bf{p}},{\bm{\psi }},{\bm{\rho }},{\bm{\theta }}}~ {R_{\rm{sum}}},\\
		s.t.\qquad &{{\bf{q}}_I} = {\bf{q}}\left[ 1 \right],\\
		&{{\bf{q}}_F} = {\bf{q}}\left[ N \right],\\
		&{\left\| {{\bf{q}}\left[ {n + 1} \right] - {\bf{q}}\left[ n \right]} \right\|^2} \le {\left( {{V_{\max }}\delta } \right)^2},n=1,...,N-1,\\
		&{p_k}\left[ n \right] \ge 0,\forall k,n,\\
		&\sum\limits_{k = 1}^K {{p_k}\left[ n \right]}  \le {P_{\max }},\forall n,\\
		&0 \le {\theta _m}\left[ n \right] \le 2\pi ,\forall m,n,\\
		&0 \le {\rho _k}\left[ n \right] \le 1,\forall k,n,\\
		&{\psi _{i,k}}\left[ n \right] = \left\{ {\begin{array}{*{20}{c}}
				{1,{\rm{ ~}}if{\rm{~}}{H_i}\left[ n \right] > {H_k}\left[ n \right]}\\
				{0,{\rm{~~~~~~~}}otherwise{\rm{  ~~~  }}}
		\end{array}} \right.,\\
  		&{\psi _{i,k}}\left[ n \right] + {\psi _{k,i}}\left[ n \right] = 1,\forall n,i \ne k,\\
  		&{p_k}\left[ n \right] \ge {\psi _{i.k}}{p_i}\left[ n \right],\forall n,i \ne k,\\
		&{E_k\left[ n \right]} \ge \chi_{th} ,\forall k,n,
	\end{align}
\end{subequations}
where (27b)-(27d) denote the UAV trajectory constraint. (27e), (27f) and (27k) are the UAV transmit power allocation constraint. (27g) denotes the IRS reflection coefficients constraint and (27h) is the PS ratio constraint in SWIPT. (27i) and (27j) are the SIC decoding order constraint in NOMA. (27l) denotes the energy harvesting threshold constraint. It can be seen that the joint optimizaton problem ${\cal P}0$ is a non-convex optimization problem since the optimization variables are highly coupled, and the objective function is not joint concave with respect to (w.r.t.) the optimization variables. In addition, the constraints (27i)-(27k) contain binary constraints. Hence, the problem ${\cal P}0$ is a mixed integer non-convex optimization problem, and it is challenging to solve the problem ${\cal P}0$ directly. Next, based on the AO algorithm framework, we propose an efficient iterative algorithm to obtain a high-quality suboptimal solution.

\newcounter{my}
\begin{figure*}[!t]
	\normalsize
	\setcounter{my}{\value{equation}}
	\setcounter{equation}{28}
	\begin{equation}
		\begin{aligned}
			\mathbb{E}\left\{ {{H_k}\left[ n \right]} \right\} = \mathbb{E}\left\{ {{{\left| {\left( {{{\mathord{\buildrel{\lower3pt\hbox{$\scriptscriptstyle\frown$}} 
											\over h} }_{{\rm{UU}},k}}\left[ n \right] + {{\mathord{\buildrel{\lower3pt\hbox{$\scriptscriptstyle\smile$}} 
											\over h} }_{{\rm{UU}},k}}\left[ n \right]} \right) + \left( {{\bf{\mathord{\buildrel{\lower3pt\hbox{$\scriptscriptstyle\frown$}} 
											\over h} }}_{{\rm{IU}},k}^H + {\bf{\mathord{\buildrel{\lower3pt\hbox{$\scriptscriptstyle\smile$}} 
											\over h} }}_{{\rm{IU}},k}^H} \right){\bf{\Theta }}\left[ n \right]{{\bf{h}}_{{\rm{UI}}}}\left[ n \right]} \right|}^2}} \right\} = {\left| {{{\mathord{\buildrel{\lower3pt\hbox{$\scriptscriptstyle\frown$}} 
								\over h} }_{{\rm{UU}},k}}\left[ n \right] + {\bf{\mathord{\buildrel{\lower3pt\hbox{$\scriptscriptstyle\frown$}} 
								\over h} }}_{{\rm{IU}},k}^H{\bf{\Theta }}\left[ n \right]{{\bf{h}}_{{\rm{UI}}}}\left[ n \right]} \right|^2} +\\
			\mathbb{E}\left\{ {{{\left| {{{\mathord{\buildrel{\lower3pt\hbox{$\scriptscriptstyle\smile$}} 
										\over h} }_{{\rm{UU}},k}}\left[ n \right]} \right|}^2}} \right\} +\mathbb{E} \left\{ {{{\left| {{\bf{\mathord{\buildrel{\lower3pt\hbox{$\scriptscriptstyle\smile$}} 
										\over h} }}_{{\rm{IU}},k}^H{\bf{\Theta }}\left[ n \right]{{\bf{h}}_{{\rm{UI}}}}\left[ n \right]} \right|}^2}} \right\} = {\left| {{{\mathord{\buildrel{\lower3pt\hbox{$\scriptscriptstyle\frown$}} 
								\over h} }_{{\rm{UU}},k}}\left[ n \right] + {\bf{\mathord{\buildrel{\lower3pt\hbox{$\scriptscriptstyle\frown$}} 
								\over h} }}_{{\rm{IU}},k}^H{\bf{\Theta }}\left[ n \right]{{\bf{h}}_{{\rm{UI}}}}\left[ n \right]} \right|^2} + \frac{{{\beta _0} - \vartheta_1 }}{{{{\left( {{d_{{\rm{UU}},k}}\left[ n \right]} \right)}^\alpha }}} + \frac{{M{\beta _0}\left( {{\beta _0} - \vartheta_2 } \right)}}{{{{\left( {{d_{{\rm{UI}}}}\left[ n \right]} \right)}^2}{{\left( {{d_{{\rm{IU}},k}}} \right)}^\gamma }}}\\
			\buildrel \Delta \over = {\xi _k}\left[ n \right],\forall k,n,
		\end{aligned}
	\end{equation}
	\setcounter{equation}{\value{my}}
	\hrulefill
	\vspace*{4pt}
\end{figure*}

\newcounter{my2}
\begin{figure*}[!t]
	\normalsize
	\setcounter{my}{\value{equation}}
	\setcounter{equation}{29}
	\begin{equation}
		\begin{aligned}
			\mathbb{E}\left\{ {{R_k}\left[ n \right]} \right\} \approx \mathbb{E}\left\{ {{{\log }_2}\left( {1 + \frac{{\left( {1 - {\rho _k}\left[ n \right]} \right){p_k}\left[ n \right]}}{{\left( {1 - {\rho _k}\left[ n \right]} \right)\sum\limits_{i \ne k}^K {{\psi _{i,k}}\left[ n \right]{p_i}\left[ n \right]}  + \frac{{\left( {1 - {\rho _k}\left[ n \right]} \right)\sigma _k^2 + \delta _k^2}}{{\mathbb{E}\left\{ {{H_k}\left[ n \right]} \right\}}}}}} \right)} \right\} = \\
			{\log _2}\left( {1 + \frac{{\left( {1 - {\rho _k}\left[ n \right]} \right){p_k}\left[ n \right]}}{{\left( {1 - {\rho _k}\left[ n \right]} \right)\sum\limits_{i \ne k}^K {{\psi _{i,k}}\left[ n \right]{p_i}\left[ n \right]}  + \frac{{\left( {1 - {\rho _k}\left[ n \right]} \right)\sigma _k^2 + \delta _k^2}}{{{\xi _k}\left[ n \right]}}}}} \right) \buildrel \Delta \over = {{\tilde R}_k}\left[ n \right], \forall n,i \ne k,
		\end{aligned}
	\end{equation}
	\setcounter{equation}{\value{my2}}
	\hrulefill
	\vspace*{4pt}
\end{figure*}

\section{Joint Optimization Algorithm for the IRS Empowered UAV SWIPT Networks}
Since the objective function of problem ${\cal P}0$ contains random variables, we take expectation on it to transform the problem. Then, for the transformed problem, we use AO algorithm to solve it. More specifically, we divide the optimization problem into four sub-problmes, i.e., $\left\{ {{\bf{Q}},{\bm{\psi }}} \right\}$, ${\bf{p}}$, ${\bm{\rho }}$ and ${\bm{\theta }}$. For given UAV transmit power allocation ${\bf{p}}$, user PS ratio ${\bm{\rho }}$ and IRS reflection coefficient ${\bm{\theta }}$, the trajectory of UAV ${\bf{Q}}$ and SIC decoding order ${\bm{\psi }}$ can be obtained. Next, for given the UAV trajectory ${\bf{Q}}$, SIC decoding order ${\bm{\psi }}$, UAV transmit power allocation ${\bf{p}}$ and user PS ratio ${\bm{\rho }}$, we can get the IRS reflection coefficient ${\bm{\theta }}$. Then, UAV transmit power allocation ${\bf{p}}$ can be obtained when the UAV trajectory ${\bf{Q}}$, SIC decoding order ${\bm{\psi }}$, IRS reflection coefficient ${\bm{\theta }}$ and user PS ratio ${\bm{\rho }}$ are fixed. Fianlly, We can also get user PS ratio ${\bm{\rho }}$ by fixing the UAV trajectory ${\bf{Q}}$, SIC decoding order ${\bm{\psi }}$, IRS reflection coefficient ${\bm{\theta }}$ and UAV transmit power allocation ${\bf{p}}$. The four sub-problems are optimized alternately until convergence is achieved, and the SCA, penalty function method and DC programming are applied when we solving the above sub-problems. Herein, we first introduce the joint UAV trajectory, SIC decoding order, UAV transmit power allocation, PS ratio and IRS reflection coefficient optimization algorithm, and then explain in detail how to solve each sub-problem. The overall optimization algorithm can be summarized as $\textbf{Algorithm 1}$.

\subsection{Optimization Problem Transformation}
Since the objective function ${R_k}\left[ n \right]$ of the optimization problem ${\cal P}0$ contains random variables, we take the expectation $\mathbb{E}\left\{ {{R_k}\left[ n \right]} \right\}$ to analyze it. Since the probability distribution of ${R_k}\left[ n \right]$ is difficult to obtain, it is not easy to find a closed-form solution of $\mathbb{E}\left\{ {{R_k}\left[ n \right]} \right\}$. Thus, we use the following $\textbf{Lemma~1}$ to approximate $\mathbb{E}\left\{ {{R_k}\left[ n \right]} \right\}$.

$\textbf{Lemma~1:}$ If $X$ is a positive independent random variable, for any $\phi>0$, $\omega>0 $ and $\varphi>0$, the following approximation result holds
\setcounter{equation}{27}
\begin{equation}
	\mathbb{E}\left\{ {{{\log }_2}\left( {1 + \frac{\phi }{{\varphi  + \frac{\omega}{X}}}} \right)} \right\} \approx \mathbb{E}\left\{ {{{\log }_2}\left( {1 + \frac{\phi }{{\varphi  + \frac{\omega}{\mathbb{E}{\left\{ X \right\}}}}}} \right)} \right\}.
\end{equation}

\emph{Proof:} The proof of $\textbf{Lemma~1}$ is similar to Theorem 1 in \cite{9121338}, which is omitted here.  $\hfill\blacksquare$

We first take an expectation of ${H_k}\left[ n \right]$ as shown in Eq. (29), with ${{\mathord{\buildrel{\lower3pt\hbox{$\scriptscriptstyle\frown$}} 
			\over h} }_{{\rm{UU}},k}}\left[ n \right] = \sqrt {\frac{{{\vartheta _1}}}{{{{\left( {{d_{{\rm{UU}},k}}\left[ n \right]} \right)}^\alpha }}}} {\rm{ }}h_{{\rm{UU}},k}^{{\rm{LoS}}}\left[ n \right] $, ${{\mathord{\buildrel{\lower3pt\hbox{$\scriptscriptstyle\smile$}} 
			\over h} }_{{\rm{UU}},k}}\left[ n \right] = \sqrt {\frac{{{\beta _0} - {\vartheta _1}}}{{{{\left( {{d_{{\rm{UU}},k}}\left[ n \right]} \right)}^\alpha }}}} h_{{\rm{UU}},k}^{{\rm{NLoS}}}\left[ n \right]$, ${{{\bf{\mathord{\buildrel{\lower3pt\hbox{$\scriptscriptstyle\frown$}} 
					\over h} }}}_{{\rm{IU}},k}} = \sqrt {\frac{{{\vartheta _2}}}{{{{\left( {{d_{{\rm{IU}},k}}} \right)}^\gamma }}}} {\bf{h}}_{{\rm{IU}},k}^{{\rm{LoS}}}$, ${{{\bf{\mathord{\buildrel{\lower3pt\hbox{$\scriptscriptstyle\smile$}} 
					\over h} }}}_{{\rm{IU}},k}} = \sqrt {\frac{{{\beta _0} - {\vartheta _2}}}{{{{\left( {{d_{{\rm{IU}},k}}} \right)}^\gamma }}}} {\bf{h}}_{{\rm{IU}},k}^{{\rm{NLoS}}}$, ${\vartheta _1} = \frac{{{\beta _0}{\kappa _1}}}{{1 + {\kappa _1}}}$ and ${\vartheta _2} = \frac{{{\beta _0}{\kappa _2}}}{{1 + {\kappa _2}}}$. According to $\textbf{Lemma~1}$, the expectation of the objective function ${R_k}\left[ n \right]$ can be approximately expressed as Eq. (30). Therefore, the objective function of problem ${\cal P}0$ can be transformed into ${\tilde R_{{\rm{sum}}}}$, which can be expressed as
\begin{algorithm}[H]
	\caption{Joint UAV Trajectory, SIC Decoding Order, UAV transmit power Allocation, PS Ratio and IRS Reflection Coefficient Optimization Algorithm} 
	\begin{algorithmic}[1]
		\State Initialize ${{\bf{Q}}^{\left( 0 \right)}}$, ${{\bm{\psi }}^{\left( 0 \right)}}$, ${{\bm{\rho }}^{\left( 0 \right)}}$, ${{\bf{p}}^{\left( 0 \right)}}$ and ${{\bm{\theta }}^{\left( 0 \right)}}$. Let $r=0$, $\varepsilon  = {10^{ - 3}}$.
		\Repeat
		\State Solve the sub-problem 1 for given ${{\bm{\rho }}^{\left( r \right)}}$, ${{\bf{p}}^{\left( r \right)}}$ and ${{\bm{\theta }}^{\left( r \right)}}$, and obtain UAV trajectory ${{\bf{Q}}^{\left( r + 1 \right)}}$ and SIC decoding order ${{\bm{\psi }}^{\left( r+1 \right)}}$.
		\State Solve the sub-problem 2 for given ${{\bf{Q}}^{\left( r \right)}}$, ${{\bm{\psi }}^{\left( r \right)}}$, ${{\bf{p}}^{\left( r \right)}}$ and ${{\bm{\theta }}^{\left( r \right)}}$, and obtain PS ratio ${{\bm{\rho }}^{\left( r + 1\right)}}$.
		\State Solve the sub-problem 3 for given ${{\bf{Q}}^{\left( r \right)}}$, ${{\bm{\psi }}^{\left( r \right)}}$, ${{\bm{\rho }}^{\left( r \right)}}$ and ${{\bm{\theta }}^{\left( r \right)}}$, and obtain UAV transmit power allocation ${{\bf{p}}^{\left( r + 1\right)}}$.
		\State Solve the sub-problem 4 for given ${{\bf{Q}}^{\left( r \right)}}$, ${{\bm{\psi }}^{\left( r \right)}}$, ${{\bm{\rho }}^{\left( r \right)}}$ and ${{\bf{p}}^{\left( r \right)}}$, and obtain IRS reflection coefficient ${{\bm{\theta }}^{\left( r + 1\right)}}$.
		\State Update $r=r+1$.
		\Until  The fractional decrease of the objective value is below a threshold $\varepsilon$.
		\State \Return UAV trajectory, SIC decoding order, UAV transmit power allocation, PS ratio and IRS reflection coefficient.
	\end{algorithmic}
\end{algorithm}
\setcounter{equation}{30}
\begin{equation}
	{\tilde R_{{\rm{sum}}}} = \sum\limits_{n = 1}^N {\sum\limits_{k = 1}^K {{{\tilde R}_k}\left[ n \right]} } .
\end{equation}

In addition, according to Eq. (14), it can be seen that the SIC decoding order of users is determined by ${{H_k}\left[ n \right]}$. For ease of analysis, we approximate Eq. (14) as
\begin{equation}
	{\psi _{i,k}}\left[ n \right] = \left\{ {\begin{array}{*{20}{c}}
			{1,{\rm{~}}if{\rm{~~}}{d_{{\rm{UU}},k}}\left[ n \right] > {d_{{\rm{UU}},i}}\left[ n \right]}\\
			{0,{\rm{ ~~~~~~~~~  }}otherwise{\rm{   ~~~~~~~~~  }}}
	\end{array}} \right..
\end{equation}
This approximation indicates that the decoding order of the users is determined by the distance between the user and the UAV, which has practical significance. First, the cascaded channels of UI-channel and IU-channel have a large path loss, so the channel power gain is largely determined by the UU-channel. Second, for UU-channel, small-scale fading is negligible compared to large-scale fading. Therefore, the channel power gain is usually determined by the distance from the UAV to the user, i.e., the shorter the distance, the higher the channel power gain.

Therefore, the optimization problem ${\cal P}0$ can be transformed into the problem ${\cal P}1$ as follows
\begin{subequations}
	\begin{align}
		{\cal P}1:~~~~&\mathop {{\rm{max}}}\limits_{{\bf{Q}},{\bf{p}},{\bm{\psi }},{\bm{\rho }},{\bm{\theta }}}~ {\tilde R_{\rm{sum}}},\\
		s.t.\qquad &\textrm {(27b)-(27h), (32), (27j)-(27l)}.
	\end{align}
\end{subequations}

\subsection{Optimization of UAV Trajectory ${\bf{Q}}$ and SIC Decoding Order ${\bm{\psi }}$}
Given UAV transmit power allocation ${\bf{p}}$, user PS ratio ${\bm{\rho }}$ and IRS reflection coefficient ${\bm{\theta }}$, the UAV trajectory ${\bf{Q}}$ and SIC decoding order ${\bm{\psi }}$ optimization problem can be given by
\begin{subequations}
	\begin{align}
		{\cal P}2:~~~~&{\mathop {{\rm{max}}}\limits_{\bf{Q},{\bm{\psi }}} \;{{\tilde R}_{{\rm{sum}}}},}\\
		s.t.\qquad &\textrm {(27b)-(27d), (27i)-(27l)}.
	\end{align}
\end{subequations}
The problem ${\cal P}2$ is still a mixed integer non-convex optimization problem due to the non-concave objective function, integer constraints (27i)-(27k) and non-convex constraint (27l). Herein, we introduce some auxiliary variables to deal with the problem ${\cal P}2$. Let $\left\{ {{u_k}\left[ n \right] > 0,\forall k, n} \right\}$ denote the upper bound of ${d_{{\rm{UU}},k}}\left[ n \right]$, and $\left\{ {u\left[ n \right] > 0,\forall n} \right\}$ denote the upper bound of ${{d_{{\rm{UI}}}}\left[ n \right]}$. Therefore, they satisfy
\begin{equation}
	{\left( {{u_k}\left[ n \right]} \right)^2} \ge {\left\| {{\bf{q}}\left[ n \right] - {{\bf{w}}_k}} \right\|^2},\forall k, n,
\end{equation}
\begin{equation}
	{\left( {u\left[ n \right]} \right)^2} \ge {\left\| {{\bf{q}}\left[ n \right] - {{\bf{w}}_r}} \right\|^2},\forall k.
\end{equation}
Therefore, the lower bound of the expected combined channel power gain, denoted by $\left\{ {{\underline{\xi} _k}\left[ n \right],\forall k, n} \right\}$, can be expressed as
\begin{equation}
	\begin{aligned}
		{\underline{\xi} _k}\left[ n \right]  =& {\beta _0}{\left( {{u_k}\left[ n \right]} \right)^{ - \alpha }} + {A_k}\left[ n \right]{\left( {u\left[ n \right]} \right)^{ - 2}} + \\
		&{B_k}\left[ n \right]{\left( {{u_k}\left[ n \right]} \right)^{ - {\alpha  \mathord{\left/
						{\vphantom {\alpha  2}} \right.
						\kern-\nulldelimiterspace} 2}}}{\left( {u\left[ n \right]} \right)^{ - 1}},\forall k,n,
	\end{aligned}
\end{equation}
${A_k}\left[ n \right] = {\beta _0}{\left| {{\bf{\mathord{\buildrel{\lower3pt\hbox{$\scriptscriptstyle\frown$}} 
					\over h} }}_{{\rm{IU}},k}^H{\bf{\Theta }}\left[ n \right]{\bf{h}}\left[ n \right]} \right|^2} + M{\beta _0}\left( {{\beta _0} - {\vartheta _2}} \right){\left( {{d_{{\rm{IU}},k}}} \right)^{ - \gamma }}$ and ${B_k}\left[ n \right] = 2{\mathop{\rm Re}\nolimits} \left\{ {\sqrt {{\vartheta _1}{\beta _0}} {\bf{\mathord{\buildrel{\lower3pt\hbox{$\scriptscriptstyle\frown$}} 
				\over h} }}_{{\rm{IU}},k}^H{\bf{\Theta }}\left[ n \right]{\bf{h}}\left[ n \right]} \right\}$, ${\mathop{\rm Re}\nolimits} \left\{  \cdot  \right\}$ is the operator for taking the real part of complex numbers. In addition, we further introduce the auxiliary variable as follows
\begin{equation}
	\begin{aligned}
		{\Lambda _k}\left[ n \right] = \left( {1 - {\rho _k}\left[ n \right]} \right)\sum\limits_{i \ne k}^K {{\psi _{i,k}}\left[ n \right]{p_i}\left[ n \right]}  + \frac{{\left( {1 - {\rho _k}\left[ n \right]} \right)\sigma _k^2 + \delta _k^2}}{{{\underline{\xi}_k}\left[ n \right]}},
	\end{aligned}	
\end{equation}
Accordingly, the objective function of problem ${\cal P}2$ can be lower bounded by	
\begin{equation}
	{{\tilde R}_k}\left[ n \right] \ge {\log _2}\left( {1 + \frac{{\left( {1 - {\rho _k}\left[ n \right]} \right){p_k}\left[ n \right]}}{{{\Lambda _k}\left[ n \right]}}} \right),\forall n,i \ne k,
\end{equation}
where the equation holds when Eq. (35) and (36) are equal respectively.	

In addition, the binary constraint (27i) can be transformed into the following constraints with continuous variables between 0 and 1,
\begin{equation}
	{\psi _{i,k}}\left[ n \right]\left( {1 - {\psi _{i,k}}\left[ n \right]} \right) \le 0,\forall n,i \ne k,
\end{equation}
\begin{equation}
	0 \le {\psi _{i,k}}\left[ n \right] \le 1,\forall n,i \ne k,
\end{equation}
\begin{equation}
	{d_{{\rm{UU}},i}}\left[ n \right] \le {\pi _i}\left[ n \right],\forall n,i,
\end{equation}
\begin{equation}
	{\psi _{i,k}}\left[ n \right]{\pi _i}\left[ n \right] \le {d_{{\rm{UU}},k}}\left[ n \right],\forall n,k \ne i,
\end{equation}
where Eq. (40) and (41) make the continuous variable ${\psi _{i,k}}\left[ n \right]$ either 0 or 1. The auxiliary variable $\left\{ {{\pi _i}\left[ n \right],\forall n,i} \right\}$ represents the upper bound of ${d_{{\rm{UU}},i}}\left[ n \right]$, and Eq. (42) and (43) ensure that when ${d_{{\rm{UU}},i}}\left[ n \right] < {d_{{\rm{UU}},k}}\left[ n \right]$, ${\psi _{i,k}}\left[ n \right] = 1$.

Therefore, the problem ${\cal P}2$ can be equivalently transformed into
\begin{subequations}
	\begin{align}
		{\cal P}2.1:~~~~&\mathop {{\rm{max}}}\limits_{{\bf{Q}},{\bm{\psi}},\Upsilon } \;\sum\limits_{n = 1}^N {\sum\limits_{k = 1}^K {\log _2}\left( {1 + \frac{{\left( {1 - {\rho _k}\left[ n \right]} \right){p_k}\left[ n \right]}}{{{\Lambda _k}\left[ n \right]}}} \right) } ,\\
		s.t.\qquad &{\underline{\xi} _k}\left[ n \right] \le {\beta _0}{\left( {{u_k}\left[ n \right]} \right)^{ - \alpha }} + {A_k}\left[ n \right]{\left( {u\left[ n \right]} \right)^{ - 2}} + \nonumber\\
		&~~~~~~{B_k}\left[ n \right]{\left( {{u_k}\left[ n \right]} \right)^{ - {\alpha  \mathord{\left/
						{\vphantom {\alpha  2}} \right.
						\kern-\nulldelimiterspace} 2}}}{\left( {u\left[ n \right]} \right)^{ - 1}},\forall k,n,\\
		&{\Lambda _k}\left[ n \right] \ge \left( {1 - {\rho _k}\left[ n \right]} \right)\sum\limits_{i \ne k}^K {{\psi _{i,k}}\left[ n \right]{p_i}\left[ n \right]}  +\nonumber\\
		&~~~~~~~~\frac{{\left( {1 - {\rho _k}\left[ n \right]} \right)\sigma _k^2 + \delta _k^2}}{{{\underline{\xi}_k}\left[ n \right]}},\forall n,i \ne k,\\
		&{\rho _k}\left[ n \right]\left( {\sum\limits_{i = 1}^K {{p_i}\left[ n \right]{\underline{\xi} _k}\left[ n \right]} \! + \!\sigma _k^2} \right)\! \ge\! {\Xi ^{ - 1}}\left( {\frac{{{\chi _{th}}}}{\delta }} \right),\forall k,n,\\
		&\textrm {(27b)-(27d), (27j), (27k), (35), (36), (40)-(43)},
	\end{align}
\end{subequations}
where ${\Xi ^{ - 1}}\left( x \right) = {b_k} - \frac{{\ln \left( {{{{\xi _k}} \mathord{\left/
					{\vphantom {{{\xi _k}} {\left( {\left( {x + {Y_k}} \right){X_k}} \right) - 1}}} \right.
					\kern-\nulldelimiterspace} {\left( {\left( {x + {Y_k}} \right){X_k}} \right) - 1}}} \right)}}{{{a_k}}}$ and $\Upsilon  = \left\{ {{u_k}\left[ n \right],u\left[ n \right],{\underline{\xi} _k}\left[ n \right],{\Lambda _k}\left[ n \right], {\pi _i}\left[ n \right], \forall n, i\ne k} \right\}$ denotes the set of all auxiliary variables. It is worth noting that at the solution to the problem ${\cal P}2.1$, if any of constraints in Eq. (35) and (36) is satisfied with strict inequality, the corresponding ${{u_k}\left[ n \right]}$ or ${{u}\left[ n \right]}$ can be decreased to make Eq. (35) and (36) satisfy with equality. Thus, the corresponding ${{\underline{\xi} _k}\left[ n \right]}$ and ${{\Lambda_k}\left[ n \right]}$ can be increased or decreased to make constraint (44b) and (44c) satisfy with equality. Therefore, at the optimal solution to problem ${\cal P}2.1$, all constraints must be satisfied with equality, i.e., the problem ${\cal P}2$ and ${\cal P}2.1$ are equivalent.
\newcounter{my3}
\begin{figure*}[!t]
	\normalsize
	\setcounter{my3}{\value{equation}}
	\setcounter{equation}{45}
	\begin{equation}
		\begin{aligned}
			f\left( {{\Lambda _k}\left[ n \right],{\psi _{i,k}}\left[ n \right]} \right) \ge \sum\limits_{n = 1}^N {\sum\limits_{k = 1}^K {\left( {{{\log }_2}\left( {1 + \frac{{\left( {1 - {\rho _k}\left[ n \right]} \right){p_k}\left[ n \right]}}{{\Lambda _k^{\left( r \right)}\left[ n \right]}}} \right) - \frac{{\left( {1 - {\rho _k}\left[ n \right]} \right){p_k}\left[ n \right]\left( {{\Lambda _k}\left[ n \right] - \Lambda _k^{\left( r \right)}\left[ n \right]} \right)}}{{\Lambda _k^{\left( r \right)}\left[ n \right]\left( {\Lambda _k^{\left( r \right)}\left[ n \right] + \left( {1 - {\rho _k}\left[ n \right]} \right){p_k}\left[ n \right]} \right)\ln 2}}} \right)} }  \\
			- {\tau _p}\sum\limits_{n = 1}^N {\sum\limits_{k = 1}^K {\sum\limits_{i \ne k}^K {\left( {{\psi _{i,k}}\left[ n \right] - {{\left( {\psi _{i,k}^{\left( r \right)}\left[ n \right]} \right)}^2} - 2\psi _{i,k}^{\left( r \right)}\left[ n \right]\left( {{\psi _{i,k}}\left[ n \right] - \psi _{i,k}^{\left( r \right)}\left[ n \right]} \right)} \right)} } }  \buildrel \Delta \over =  f{\left( {{\Lambda _k}\left[ n \right],{\psi _{i,k}}\left[ n \right]} \right)^{lb}},
		\end{aligned}
	\end{equation}
	\setcounter{equation}{\value{my3}}
\hrulefill
\vspace*{4pt}
\end{figure*}
Next, we adopt a penalty function-based approach to solve the problem ${\cal P}2.1$. By adding Eq. (40) as a penalty term to the objective function, the problem ${\cal P}2.1$ can be transformed into the problem ${\cal P}2.2$ as follows 
\begin{subequations}
	\begin{align}
		{\cal P}2.2:~&\mathop {{\rm{max}}}\limits_{{\bf{Q}},{\bm{\psi}},\Upsilon } \;\;\sum\limits_{n = 1}^N {\sum\limits_{k = 1}^K {{{\log }_2}} \left( {1 + \frac{{\left( {1 - {\rho _k}\left[ n \right]} \right){p_k}\left[ n \right]}}{{{\Lambda _k}\left[ n \right]}}} \right)}\nonumber\\
		&~~~~~~~~~~~~-{\tau _p}\sum\limits_{n = 1}^N {\sum\limits_{k = 1}^K {\sum\limits_{i \ne k}^K {\left( {{\psi _{i,k}}\left[ n \right]\left( {1 - {\psi _{i,k}}\left[ n \right]} \right)} \right)} } }  ,\\
		s.t.\qquad &\textrm {(27b)-(27d), (27j), (27k), (35), (36),}\nonumber\\
		& \textrm {(41)-(43), (44b)-(44d)},
	\end{align}
\end{subequations}				
where ${\tau _p}>0$ denotes the penalty factor, and its role is to penalize the objective function when $\psi _{i,k}$ belongs to 0 to 1. It can be seen that when ${\tau _p} \to \infty $, the problem ${\cal P}2.1$ and the problem ${\cal P}2.2$ are equivalent. However, the problem ${\cal P}2.2$ is still non-convex optimization problem due to the non-concave objective function and non-convex constraints (35), (36), (43) and (44b). Next, we apply SCA to iteratively obtain a suboptimal solution to problem ${\cal P}2.2$. 

We define that $f\left( {{\Lambda_k}\left[ n \right]},{{\psi _{i,k}}\left[ n \right]} \right)$ denotes the objective function of the problem ${\cal P}2.2$, which is convex w.r.t ${{B_k}\left[ n \right]}$ and ${{\psi _{i,k}}\left[ n \right]}$. For the $r$-th iteration of SCA, the lower bound of $f\left( {{\Lambda_k}\left[ n \right]},{{\psi _{i,k}}\left[ n \right]} \right)$ can be given by Eq. (46), where ${{B_k}{{\left[ n \right]}^{\left( r \right)}}}$ and $\psi _{i,k}^{\left( r \right)}\left[ n \right]$ are value of the $r$-th iteration. Similarly, we can apply SCA to carry out the first-order Taylor expansion of the left-hand-side (LHS) convex functions of constraints (35) and (36), and obtain their lower bounds respectively as following
\setcounter{equation}{46}
\begin{equation}
	\begin{aligned}
		{\left( {u_k^{\left( r \right)}\left[ n \right]} \right)^2} + 2u_k^{\left( r \right)}\left[ n \right]\left( {{u_k}\left[ n \right] - u_k^{\left( r \right)}\left[ n \right]} \right)\\
		 \ge {\left\| {{\bf{q}}\left[ n \right] - {{\bf{w}}_k}} \right\|^2},\forall k,n,
	\end{aligned}
\end{equation}
and
\begin{equation}
	{\left( {{u^{\left( r \right)}}\left[ n \right]} \right)^2} + 2{u^{\left( r \right)}}\left[ n \right]\left( {u\left[ n \right] - {u^{\left( r \right)}}\left[ n \right]} \right) \ge {\left\| {{\bf{q}}\left[ n \right] - {{\bf{w}}_r}} \right\|^2},\forall n.
\end{equation}
Moreover, the non-convex constraint (43) can be rewritten as follows
\begin{equation}
	\begin{aligned}
		\frac{{{{\left( {{\psi _{i,k}}\left[ n \right] + {\pi _i}\left[ n \right]} \right)}^2}}}{4} - \frac{{{{\left( {{\psi _{i,k}}\left[ n \right] - {\pi _i}\left[ n \right]} \right)}^2}}}{4} \le \\{\left\| {{\bf{q}}\left[ n \right] - {{\bf{w}}_k}} \right\|^2},\forall n,k \ne i.
	\end{aligned}
\end{equation}
It can be seen that the LHS of the Eq. (49) is the difference of the two convex functions w.r.t ${{\psi _{i,k}}\left[ n \right]}$ and ${\pi _i}\left[ n \right]$, and the right-hand-side (RHS) is also a convex function w.r.t ${{\bf{q}}\left[ n \right]}$. Hence, for the $r$-th SCA iteration, Eq. (49) can be approximately expressed as
\begin{equation}
	\begin{aligned}
		d_{i,k}^{ub}\left[ n \right] \le {\left\| {{{\bf{q}}^{\left( r \right)}}\left[ n \right] - {{\bf{w}}_k}} \right\|^2} + 2{\left( {{\bf{q}}\left[ n \right] - {{\bf{w}}_k}} \right)^T}\\\left( {{\bf{q}}\left[ n \right] - {{\bf{q}}^{\left( r \right)}}\left[ n \right]} \right),\forall n,k \ne i,
	\end{aligned}
\end{equation}
where $d_{i,k}^{ub}\left[ n \right] = \frac{{{{\left( {{\psi _{i,k}}\left[ n \right] + {\pi _i}\left[ n \right]} \right)}^2}}}{4} + \frac{{{{\left( {\psi _{i,k}^{\left( r \right)}\left[ n \right] - \pi _i^k\left[ n \right]} \right)}^2} - 2\left( {\psi _{i,k}^{\left( r \right)}\left[ n \right] - \pi _i^k\left[ n \right]} \right)\left( {{\psi _{i,k}}\left[ n \right] - {\pi _i}\left[ n \right]} \right)}}{4},\forall n,k \ne i$.
In addition, the RHS of constraint (44b) is intractable due to the AoA in ${A_k}\left[ n \right]$ and ${B_k}\left[ n \right]$ depend on the UAV location in the $n$-th time slot ${{\bf{q}}\left[ n \right]}$. Herein, we introduce the following constraint
\begin{equation}
	{\left\| {{\bf{q}}\left[ n \right] - {{\bf{q}}^{\left( r \right)}}\left[ n \right]} \right\|^2} \le \delta _{\max }^2,\forall n,
\end{equation}
where ${{{\bf{q}}^{\left( r \right)}}\left[ n \right]}$ is the value of the $r$-th SCA iteration, and $\delta _{\max }$ denotes the maximum allowed displacement of UAV after each SCA iteration. When the value of $\delta _{\max }$ is small enough, we can consider that AoA is approximately unchanged after each SCA iteration. Therefore, ${A_k}\left[ n \right]$ and ${B_k}\left[ n \right]$ also remain unchanged. The UAV trajectory optimization of the $(r+1)$-th SCA iteration is based on the AoA obtained at the $r$-th iteration. In order to ensure the accuracy of the approximation, according to \cite{8247211}, we require that the ratio of the maximum allowed deployment of the $n$-th time slot to the minimum height of the UAV should meet ${{{\delta _{\max }}} \mathord{\left/
		{\vphantom {{{\delta _{\max }}} {{{\left( {{h_u}} \right)}_{\min }}}}} \right.
		\kern-\nulldelimiterspace} {{{\left( {{h_u}} \right)}_{\min }}}} \le {\varepsilon _{\max }}$, i.e., the value of $\delta _{\max }$ under the accuracy threshold ${\varepsilon _{\max }}$ can be represented by ${\delta _{\max }} \le {\left( {{h_u}} \right)_{\min }}{\varepsilon _{\max }}$\footnote{It is worth noting that a sufficiently small ${\varepsilon _{\max }}$ will improve the accuracy of the approximation, but it will also increase the computational complexity. Therefore, the choice of a suitable ${\varepsilon _{\max }}$ can well balance the relationship between accuracy and complexity.}. Accordingly, the RHS of constraint (44b) only depends on ${{u_k}\left[ n \right]}$ and ${{u}\left[ n \right]}$. We use the following lemma to deal with constraint (44b).

$\textbf{Lemma~2:}$ For given ${a_1} > 0$, ${a_2} > 0$ and ${a_3} > 0$, $g_1\left( {{x_1},{x_2}} \right) = {a_1}{\left( {{x_1}} \right)^{ - \alpha }} + {a_2}{\left( {{x_2}} \right)^{ - 2}}$ and ${g_2}\left( {{x_1},{x_2}} \right) = {a_3}{\left( {{x_1}} \right)^{{{ - \alpha } \mathord{\left/
				{\vphantom {{ - \alpha } 2}} \right.
				\kern-\nulldelimiterspace} 2}}}{\left( {{x_2}} \right)^{ - 1}}$ are both convex jointly w.r.t. ${x_1} > 0$ and ${x_2} > 0$.

\emph{Proof:} By proving that when $x_1>0$ and $x_2>0$, the Hessian matrices of ${g_1}\left( {{x_1},{x_2}} \right)$ and ${g_2}\left( {{x_1},{x_2}} \right)$ are positive semi-definite, therefore, both are convex functions. The proof of \textbf{Lemma 2} is completed.  $\hfill\blacksquare$

Let $\tilde g = {\beta _0}{\left( {{u_k}\left[ n \right]} \right)^{ - \alpha }} + {A_k}\left[ n \right]{\left( {u\left[ n \right]} \right)^{ - 2}}$ and $\tilde h = {\left( {{u_k}\left[ n \right]} \right)^{{{ - \alpha } \mathord{\left/
				{\vphantom {{ - \alpha } 2}} \right.
				\kern-\nulldelimiterspace} 2}}}{\left( {u\left[ n \right]} \right)^{ - 1}}$. The RHS of the Eq. (44b) can be given by $\tilde g + {B_k}\left[ n \right]\tilde h$. Since ${\beta _0} > 0$, if ${B_k}>0$, then $\tilde g + \left| {{B_k}\left[ n \right]} \right|\tilde h$ is convex jointly w.r.t. ${{u_k}\left[ n \right]}$ and ${u\left[ n \right]}$ according to $\textbf{Lemma~2}$. Otherwise, $\tilde g - \left| {{B_k}\left[ n \right]} \right|\tilde h$ is the difference between the two convex functions. Thus, constraint (44b) is non-convex. Given the value of the $r$-th iteration ${u_k^{\left( r \right)}\left[ n \right]}$ and ${{u^{\left( r \right)}}\left[ n \right]}$, we apply SCA to obtain the lower bound of $\tilde g + {B_k}\left[ n \right]\tilde h$, which can be given by
\begin{equation}
	{\left( {\tilde g + {B_k}\left[ n \right]\tilde h} \right)^{lb}} = \left\{ {\begin{array}{*{20}{c}}
			{{{\tilde g}^{lb}} + \left| {{B_k}\left[ n \right]} \right|{{\tilde h}^{lb}}}\\
			{{{\tilde g}^{lb}} - \left| {{B_k}\left[ n \right]} \right|\tilde h}
	\end{array}} \right..
\end{equation}
where ${{{\tilde g}^{lb}}}$ and ${{{\tilde h}^{lb}}}$ are as follows. 
\begin{equation}
	\begin{aligned}
		{{\tilde g}^{lb}} = {\beta _0}{\left( {u_k^{\left( r \right)}\left[ n \right]} \right)^{ - \alpha }} \!-\! \alpha {\left( {u_k^{\left( r \right)}\left[ n \right]} \right)^{ - \alpha  - 1}}\left( {{u_k}\left[ n \right] - u_k^{\left( r \right)}\left[ n \right]} \right) \\+ {A_k}\left[ n \right]{\left( {{u^{\left( r \right)}}\left[ n \right]} \right)^{ - 2}} - 2{A_k}\left[ n \right]{\left( {{u^{\left( r \right)}}\left[ n \right]} \right)^{ - 3}}\\\left( {u\left[ n \right] - {u^{\left( r \right)}}\left[ n \right]} \right),\forall k,n,
	\end{aligned}
\end{equation}
and
\begin{equation}
	\begin{aligned}
		{{\tilde h}^{lb}} \!= \!{\left( {u_k^{\left( r \right)}\left[ n \right]} \right)^{{{ - \alpha } \mathord{\left/
						{\vphantom {{ - \alpha } 2}} \right.
						\kern-\nulldelimiterspace} 2}}}{\left( {{u^{\left( r \right)}}\left[ n \right]} \right)^{ - 1}} \!-\! \frac{\alpha }{2}{\left( {u_k^{\left( r \right)}\left[ n \right]} \right)^{{{ - \alpha } \mathord{\left/
						{\vphantom {{ - \alpha } {2 - 1}}} \right.
						\kern-\nulldelimiterspace} {2 - 1}}}}\\{\left(\! {{u^{\left( r \right)}}\left[ n \right]} \right)^{ - 1}}\left( {{u_k}\left[ n \right] \!- \!u_k^{\left( r \right)}\left[ n \right]} \right)\! -\! {\left( {u_k^{\left( r \right)}\left[ n \right]} \right)^{{{ - \alpha } \mathord{\left/
						{\vphantom {{ - \alpha } 2}} \right.
						\kern-\nulldelimiterspace} 2}}}{\left( {{u^{\left( r \right)}}\left[ n \right]} \right)^{ - 2}}\\\left( {u\left[ n \right] - {u^{\left( r \right)}}\left[ n \right]} \right),\forall k,n,
	\end{aligned}
\end{equation}
Therefore, constraint (44b) can be transformed into
\begin{equation}
	{\underline{\xi} _k}\left[ n \right] \le {\left( {\tilde g + {B_k}\left[ n \right]\tilde h} \right)^{lb}},\forall k,n.
\end{equation}
Accordingly, the problem ${\cal P}2.2$ can be transformed into the problem ${\cal P}2.3$, which can be expressed as
\begin{subequations}
	\begin{align}
		{\cal P}2.3:~~~~&\mathop {{\rm{max}}}\limits_{{\bf{Q}},\bm{\psi},\Upsilon }~f{\left( {{\Lambda _k}\left[ n \right],{\psi _{i,k}}\left[ n \right]} \right)^{lb}} ,\\
		s.t.\qquad &\textrm {(27b)-(27d), (27j), (27k), (41), (42),}\nonumber \\
			&\textrm {(44c), (44d), (47), (48), (50), (55).}
	\end{align}
\end{subequations}
The problem ${\cal P}2.2$ is a standard convex optimization problem, which can be solved by using the standard solvers, e.g., CVX toolbox \cite{grant2014cvx}.
\subsection{Optimization of PS Ratio ${\bm{\rho }}$}
For given UAV trajectory ${\bf{Q}}$, SIC decoding order $\bm{\psi}$, UAV transmit power allocation ${\bf{p}}$, and IRS reflection coefficient ${\bm{\theta }}$, the user PS ratio ${\bm{\rho }}$ optimization problem can be transformed into the problem $ {\cal P}3$, which can be expressed as
\begin{subequations}
	\begin{align}
		{\cal P}3:~~~~&\mathop {{\rm{max}}}\limits_{{\bm{\rho}}}~\sum\limits_{n = 1}^N {\sum\limits_{k = 1}^K {{{\tilde R}_k}\left[ n \right]} } ,\\
		s.t.\qquad &\textrm {(27h)},\\
		&{\rho _k}\left[ n \right]\left( {\sum\limits_{i = 1}^K {{p_i}\left[ n \right]{\xi _k}\left[ n \right]} \! + \!\sigma _k^2} \right)\! \ge\! {\Xi ^{ - 1}}\left( {\frac{{{\chi _{th}}}}{\delta }} \right),\forall k,n.
	\end{align}
\end{subequations}
It can be proved that the objective function ${{{\tilde R}_k}\left[ n \right]}$ is concave w.r.t $\rho_k\left[n\right]$, and it is omitted here. Hence, the problem ${\cal P}3$ is a standard convex optimization problem, which can be solved by using CVX toolbox \cite{grant2014cvx}.
\subsection{Optimization of UAV Transmit Power Allocation ${\bf{p}}$}
For given UAV trajectory ${\bf{Q}}$, SIC decoding order $\bm{\psi}$, user PS ratio ${\bm{\rho }}$, and IRS reflection coefficient ${\bm{\theta }}$, the  UAV transmit power allocation ${\bf{p}}$ optimization problem can be expressed as
\begin{subequations}
	\begin{align}
		{\cal P}4:~~~~&\mathop {{\rm{max}}}\limits_{{\bf{p}}}~\sum\limits_{n = 1}^N {\sum\limits_{k = 1}^K {{{\tilde R}_k}\left[ n \right]} } ,\\
		s.t.\qquad &\textrm {(27e), (27f), (27k),}\\
		&{\rho _k}\left[ n \right]\left( {\sum\limits_{i = 1}^K {{p_i}\left[ n \right]{\xi _k}\left[ n \right]} \! + \!\sigma _k^2} \right)\! \ge\! {\Xi ^{ - 1}}\left( {\frac{{{\chi _{th}}}}{\delta }} \right),\forall k,n.
	\end{align}
\end{subequations}
The problem ${\cal P}4$ is a non-convex optimization problem due to the non-concave objective function. The objective function of the problem ${\cal P}4$ can be further expressed as
\begin{equation}
	\begin{aligned}
		{{\tilde R}_k}\left[ n \right] \!= \! \underbrace {{{\log }_2} \!\left( \! {\left( {1 \! - \! {\rho _k}\left[ n \right]} \right)\left(  \!{\sum\limits_{i = 1}^K {{\psi _{i,k}} \!\left[ n \right]{p_i} \!\left[ n \right]{\xi _k} \!\left[ n \right]}  \! + \! \sigma _k^2} \right) \! +  \!\delta _k^2} \right)}_{{{\bar \ell }_{i,k}}\left[ n \right]} \\- \underbrace {{{\log }_2}\left( {\left( {1 - {\rho _k}\left[ n \right]} \right)\left( {\sum\limits_{i \ne k}^K {{\psi _{i,k}}\left[ n \right]{p_i}\left[ n \right]{\xi _k}\left[ n \right]}  + \sigma _k^2} \right) + \delta _k^2} \right)}_{{{\tilde \ell }_{i,k}}\left[ n \right]},\\\forall n,k \ne i.
	\end{aligned}
\end{equation}
For the convenience of analysis, we set ${\psi _{i,i}}\left[ n \right] = 1$. ${{\tilde R}_k}\left[ n \right]$ is non-concave w.r.t. ${{p_i}\left[ n \right]}$ due to it is a form of difference of concave functions. Thus, we apply SCA to obtain the upper bound of ${{{\tilde \ell }_{i,k}}\left[ n \right]}$ on the RHS of Eq. (59) as follows
\begin{equation}
	\begin{aligned}
		&{{\tilde \ell }_{i,k}}\left[ n \right] \le {{\tilde \ell }_{i,k}}\left[ n \right]\left( {p_i^{\left( r \right)}\left[ n \right]} \right) + \\
		&\frac{{\sum\limits_{i \ne k}^K {{\psi _{i,k}}\left[ n \right]{\xi _k}\left[ n \right]\left( {{p_i}\left[ n \right] - p_i^{\left( r \right)}\left[ n \right]} \right)} }}{{\left( {\left( {1 - {\rho _k}\left[ n \right]} \right)\left( {\sum\limits_{i \ne k}^K {{\psi _{i,k}}\left[ n \right]p_i^{\left( r \right)}\left[ n \right]{\xi _k}\left[ n \right]}  + \sigma _k^2} \right) + \delta _k^2} \right)\ln 2}}\\& \buildrel \Delta \over = {\left( {{{\tilde \ell }_{i,k}}\left[ n \right]} \right)^{ub}},\forall n,k \ne i,
	\end{aligned}	
\end{equation}
where ${p_i^{\left( r \right)}\left[ n \right]}$ is value of the $r$-th SCA iteration. Accordingly, the problem ${\cal P}4$ can be transformed into the problem ${\cal P}4.1$, which can be given by
\begin{subequations}
	\begin{align}
		{\cal P}4.1:~~~~&\mathop {\max }\limits_{\bf{p}} {\rm{~}}\sum\limits_{n = 1}^N {\sum\limits_{k = 1}^K {\left( {{{\bar \ell }_{i,k}}\left[ n \right] - {{\left( {{{\tilde \ell }_{i,k}}\left[ n \right]} \right)}^{ub}}} \right)} } ,\\
		s.t.\qquad &\textrm {(27e), (27f), (27k),}\\
		&{\rho _k}\left[ n \right]\left( {\sum\limits_{i = 1}^K {{p_i}\left[ n \right]{\xi _k}\left[ n \right]} \! + \!\sigma _k^2} \right)\! \ge\! {\Xi ^{ - 1}}\left( {\frac{{{\chi _{th}}}}{\delta }} \right),\forall k,n.
	\end{align}
\end{subequations}
It can be seen that the problem ${\cal P}4.1$ is a standard convex optimization problem, which can be solved by applying CVX toolbox \cite{grant2014cvx}.
\newcounter{my4}
\begin{figure*}[!t]
	\normalsize
	\setcounter{my4}{\value{equation}}
	\setcounter{equation}{64}
	\begin{equation}
		\begin{aligned}
			{{\tilde R}_k}\left[ n \right] =& \underbrace {{{\log }_2}\left( {\left( {1 - {\rho _k}\left[ n \right]} \right)\sum\limits_{i = 1}^K {{\psi _{i,k}}\left[ n \right]{p_i}\left[ n \right]{\rm{tr}}\left( {{{\bf{A}}_k}\left[ n \right]{\bf{B}}\left[ n \right]} \right) + {{\bar \upsilon }_k}\left[ n \right]} } \right)}_{{{\bar \iota }_{i,k}}\left[ n \right]} - \\
			&\underbrace {{{\log }_2}\left( {\left( {1 - {\rho _k}\left[ n \right]} \right)\sum\limits_{i \ne k}^K {{\psi _{i,k}}\left[ n \right]{p_i}\left[ n \right]{\rm{tr}}\left( {{{\bf{A}}_k}\left[ n \right]{\bf{B}}\left[ n \right]} \right) + {{\tilde \upsilon }_k}\left[ n \right]} } \right)}_{{{\tilde \iota }_{i,k}}\left[ n \right]},\forall n,k \ne i,
		\end{aligned}
	\end{equation}
	\setcounter{equation}{\value{my4}}
	\hrulefill
	\vspace*{4pt}
\end{figure*}
\newcounter{my5}
\begin{figure*}[!t]
	\normalsize
	\setcounter{my5}{\value{equation}}
	\setcounter{equation}{68}
	\begin{equation}
		\begin{aligned}
			{\nabla _{{{\bf{B}}^{\left( r \right)}}\left[ n \right]}}{{\tilde \iota }_{i,k}}\left[ n \right]\left( {{{\bf{B}}^{\left( r \right)}}\left[ n \right]} \right) = \frac{{\left( {1 - {\rho _k}\left[ n \right]} \right){\bf{A}}_k^H\left[ n \right]\sum\limits_{i \ne k}^K {{\psi _{i,k}}\left[ n \right]{p_i}\left[ n \right]} }}{{\left( {\left( {1 - {\rho _k}\left[ n \right]} \right)\sum\limits_{i \ne k}^K {{\psi _{i,k}}\left[ n \right]{p_i}\left[ n \right]{\rm{tr}}\left( {{{\bf{A}}_k}\left[ n \right]{{\bf{B}}^{\left( r \right)}}\left[ n \right]} \right) + {{\tilde \upsilon }_k}\left[ n \right]} } \right)\ln 2}},\forall n,k \ne i.
		\end{aligned}
	\end{equation}
	\setcounter{equation}{\value{my5}}
	\hrulefill
	\vspace*{4pt}
\end{figure*}
\subsection{Optimization of IRS Reflection Coefficient ${\bm{\theta }}$}
For given UAV trajectory ${\bf{Q}}$, user PS ratio ${\bm{\rho }}$, and UAV transmit power allocation ${\bf{p}}$, the   IRS reflection coefficient ${\bm{\theta }}$ optimization problem can be expressed as
\begin{subequations}
	\begin{align}
		{\cal P}5:~~~~&{\mathop {{\rm{max}}}\limits_{\bm{\theta}} {\rm{~~}}{{\tilde R}_{{\rm{sum}}}}},\\
		s.t.\qquad &\textrm {(27g), (27l)}.
	\end{align}
\end{subequations}
The problem ${\cal P}5$ is non-convex due to the non-concave objective function and non-convex constraint (27g) and (27l). We introduce auxiliary variables ${\bf{a}}_k^H\left[ n \right] = {\bf{\mathord{\buildrel{\lower3pt\hbox{$\scriptscriptstyle\frown$}} 
			\over h} }}_{{\rm{IU}},k}^H{\rm{diag}}\left( {{{\bf{h}}_{{\rm{UI}}}}\left[ n \right]} \right) \in {\mathbb{C}^{1 \times M}},\forall k,n$ and ${\bf{b}}\left[ n \right] = {\left[ {{e^{j{\theta _1}\left[ n \right]}},...,{e^{j{\theta _M}\left[ n \right]}}} \right]^T} \in {\mathbb{C}^{M \times 1}}$. Thus, the Eq. (25) can be written as follows
\begin{equation}
	{\xi _k}\left[ n \right] = {\varpi _k}\left[ n \right] + {\left| {{\bf{a}}_k^H\left[ n \right]{\bf{b}}\left[ n \right]} \right|^2},\forall k,n,
\end{equation}
where ${\varpi _k}\left[ n \right] = \frac{{{\beta _0} - {\vartheta _1}}}{{{{\left( {{d_{{\rm{UU}},k}}\left[ n \right]} \right)}^\alpha }}} + \frac{{M{\beta _0}\left( {{\beta _0} - {\vartheta _2}} \right)}}{{{{\left( {{d_{{\rm{UI}}}}\left[ n \right]} \right)}^2}{{\left( {{d_{{\rm{IU}},k}}} \right)}^\gamma }}}$. Let ${{\bf{A}}_k}\left[ n \right] = {{\bf{a}}_k}\left[ n \right]{\bf{a}}_k^H\left[ n \right] \in {\mathbb{C}^{M \times M}}$ and ${\bf{B}}\left[ n \right] = {\bf{b}}\left[ n \right]{\bf{b}}{\left[ n \right]^H} \in {\mathbb{C}^{M \times M}}$. They satisfy ${\rm{rank}}\left( {{{\bf{A}}_k}\left[ n \right]} \right) = 1$ and ${\rm{rank}}\left( {\bf{B}} \left[ n \right]\right) = 1$. Then Eq. (63) can be further expressed as
\begin{equation}
	{\xi _k}\left[ n \right] = {\varpi _k}\left[ n \right] + {\rm{tr}}\left( {{{\bf{A}}_k}\left[ n \right]{\bf{B}}}\left[ n \right] \right).
\end{equation}
Hence, the Eq. (30) can be rewritten as the Eq. (65). Next, we can apply DC programming to transform the non-convex constraint ${\rm{rank}}\left( {\bf{B}}\left[ n \right] \right) = 1$.

$\textbf{Proposition 1:}$ For the positive semi-definite matrix ${\bf{M}} \in {\mathbb{C}^{N \times N}}$, ${\rm{tr}}\left( {\bf{M}} \right) > 0$, the rank-one constraint can be expressed as the difference between two convex functions, i.e.,
\setcounter{equation}{65}
\begin{equation}
	{\rm{rank}}\left( {\bf{M}} \right) = 1 \Leftrightarrow {\rm{tr}}\left( {\bf{M}} \right) - {\left\| {\bf{M}} \right\|_2}=0,
\end{equation}
where ${\rm{tr}}\left( {\bf{M}} \right) = \sum\limits_{n = 1}^N {{\sigma _n}\left( {\bf{M}} \right)} $, ${\left\| {\bf{M}} \right\|_2} = {\sigma _1}\left( {\bf{M}} \right)$ is spectral norm, and ${\sigma _n}\left( {\bf{M}} \right)$ represents the $n$-th largest singular value of matrix ${\bf{M}}$ \cite{9352968}. 

According to $\textbf{Proposition 1}$, we transform the non-convex rank-one constraint on matrix ${\bf{B}}{\left[ n \right]}$, and then add it as a penalty term to the objective function of problem ${\cal P}5$. Therefore, problem ${\cal P}5$ can be transformed into
\begin{subequations}
	\begin{align}
		{\cal P}5.1:\;&\mathop {{\rm{max}}}\limits_{{\bf{B}}\left[ n \right]} \!\!\!\!\;\;\sum\limits_{n = 1}^N \!{\sum\limits_{k = 1}^K \!{\left( {{{\bar \iota }_{i,k}}\!\left[ n \right]\!\! -\! {{\tilde \iota }_{i,k}}\!\left[ n \right]} \right)} }\!\!  -\! {\varsigma _p}\!\left( {{\rm{tr}}\!\left( {\bf{B}}\!\left[ n \right] \right)\!\! -\! {{\left\| {\bf{B}}\!\left[ n \right] \right\|}_2}} \right),\\
		s.t.\qquad &{\rho _k}\left[ n \right]\left( {\sum\limits_{i = 1}^K  {p_i}\left[ n \right]\left( {{\varpi _k}\left[ n \right] + {\rm{tr}}\left( {{{\bf{A}}_k}\left[ n \right]{\bf{B}}\left[ n \right]} \right)} \right) + \sigma _k^2} \right) \nonumber\\
		&\ge {\Xi ^{ - 1}}\left( {\frac{{{\chi _{th}}}}{\delta }} \right),\forall k,n,\\
		&{{{\bf{B}}\left[ n \right]_{m,m}} = 1},\forall m,n,\\
		&{\bf{B}} \left[ n \right]\succeq 0,\forall n,
	\end{align}
\end{subequations}
where $\varsigma_p  > 0$ denotes the penalty factor related to the rank-one constraint. It can be seen that when ${\varsigma_p} \to \infty $, the problem ${\cal P}5.1$ and the problem ${\cal P}5$ are equivalent. The problem ${\cal P}5.1$ is still non-convex optimization problem due to the objective function is non-concave. As ${{{\tilde \iota }_{i,k}}\left[ n \right]}$ is concave w.r.t ${{\bf{B}}\left[ n \right]}$, the upper bound can be obtained by adopting SCA as follows
\begin{equation}
	\begin{aligned}
		{{\tilde \iota }_{i,k}}\left[ n \right]\! \le\! {{\tilde \iota }_{i,k}}\left[ n \right]\!\left( {{{\bf{B}}^{\left( r \right)}}\left[ n \right]} \right) \!+ \!{\rm{tr}}\left(\! {{{\left(\! {{\nabla _{{{\bf{B}}^{\left( r \right)}}\!\left[ n \right]}}{{\tilde \iota }_{i,k}}\left[ n \right]\left(\! {{{\bf{B}}^{\left( r \right)}}\!\left[ n \right]} \right)} \right)}^H}}\right.\\ \left.{\left( {{\bf{B}}\left[ n \right] - {{\bf{B}}^{\left( r \right)}}\left[ n \right]} \right)} \right) \buildrel \Delta \over = {\left( {{{\tilde \iota }_{i,k}}\left[ n \right]} \right)^{ub}},\forall n,k \ne i,
	\end{aligned}
\end{equation}
where ${\nabla _{{{\bf{B}}^{\left( r \right)}}\left[ n \right]}}{{\tilde \iota }_{i,k}}\left[ n \right]\left( {{{\bf{B}}^{\left( r \right)}}\left[ n \right]} \right)$ is denoted by Eq. (69). In addition, since ${{{\left\| {{\bf{B}}\left[ n \right]} \right\|}_2}}$ is a convex function, we can also adopt SCA to obtain its lower bound as follows
\setcounter{equation}{69}
\begin{equation}
	\begin{aligned}
		{\left\| {{\bf{B}}\left[ n \right]} \right\|_2} \!\ge\! {\left\| {{{\bf{B}}^{\left( r \right)}}\left[ n \right]} \right\|_2}\! +\! {\rm{tr}}\left(\!{ {{\bf{u}}_{\max }}\!\left( {{{\bf{B}}^{\left( r \right)}}\!\left[ n \right]} \right)\!{{\bf{u}}_{\max }}{{\left(\! {{{\bf{B}}^{\left( r \right)}}\left[ n \right]} \right)}^H}}\right.\\
		\left.{\left( {{\bf{B}}\left[ n \right] - {{\bf{B}}^{\left( r \right)}}\left[ n \right]} \right)} \right) \buildrel \Delta \over = {\left( {{{\left\| {{\bf{B}}\left[ n \right]} \right\|}_2}} \right)^{lb}},\forall n,
	\end{aligned}
\end{equation}
where ${{{\bf{u}}_{\max }}\left( {{{\bf{B}}^{\left( r \right)}}\left[ n \right]} \right)}$ denotes the eigenvector corresponding to the largest singular value of the matrix ${{{\bf{B}}^{\left( r \right)}}\left[ n \right]}$. Therefore, the non-convex problem ${\cal P}5.1$ can be approximately transformed into
\begin{subequations}
	\begin{align}
		{\cal P}5.2:\;&\!\!\sum\limits_{n = 1}^N \!{\sum\limits_{k = 1}^K \!{\left( \!{{{\bar \iota }_{i,k}}\left[ n \right] \!\!-\! {{\left( {{{\tilde \iota }_{i,k}}\left[ n \right]} \right)}^{ub}}}\! \right)} } \!\! - \!{\varsigma _p}\!\left( {{\rm{tr}}\left( {{\bf{B}}\left[ n \right]} \right)\!\! -\! {{\left( {{{\left\| {{\bf{B}}\left[ n \right]} \right\|}_2}} \right)}^{lb}}} \right),\\
		s.t.\qquad&\textrm {(67b)-(67d).}
	\end{align}
\end{subequations}
It can be seen that the problem ${\cal P}5.2$ is a standard SDP problem, which can be solved by using CVX toolbox \cite{grant2014cvx}.

\subsection{Computational Complexity and Convergence Analysis}
\subsubsection{Computational complexity analysis}
In each iteration, the problem ${\cal P}2.3$ is solved with the computational complexity of ${\cal O}\left( N ^{3.5}+KN\right)$, the problem ${\cal P}3$ and problem ${\cal P}4.1$ both are solved with computational complexity of ${\cal O}\left( KN ^{3.5}\right)$. The problem ${\cal P}5.2$ solves a SDP problem by interior point method, so the computational complexity can be represented by ${\cal O}\left( {{ M ^{3.5}}} \right)$ \cite{boyd2004convex}. We assume that the number of iterations required for the algorithm to reach convergence is $r$, the computational complexity of the proposed algorithm can be expressed as ${\cal O}\left( {r\left( {N^{3.5} + KN +\left(KN\right) ^{3.5}+ {{M}^{3.5}} } \right)} \right)$.

\subsubsection{Convergence analysis}
The convergence of the proposed joint UAV trajectory, SIC decoding order, UAV transmit power allocation, PS ratio and IRS reflection coefficient optimization in IRS empowered UAV SWIPT networks can be elaborated as follows. 

We define ${{\bf{Q}}^{\left( r \right)}}$, ${{\bm{\psi }}^{\left( r \right)}}$, ${{\bm{\rho }}^{\left( r \right)}}$, ${{\bf{p}}^{\left( r \right)}}$ and ${{\bm{\theta }}^{\left( r \right)}}$ as the $r$-th iteration solution of the problem ${\cal P}2.3$, ${\cal P}3$, ${\cal P}4.1$ and ${\cal P}5.2$. Herein, the objective function is denoted by $\Re \left( {{{\bf{Q}}^{\left( r \right)}},{{\bm{\psi }}^{\left( r \right)}},{{\bm{\rho }}^{\left( r \right)}},{{\bf{p}}^{\left( r \right)}},{{\bm{\theta }}^{\left( r \right)}}} \right)$. In the step 3 of $\textbf{Algorithm 1}$, since the UAV trajectory and SIC decoding order can be obtained for given ${{\bm{\rho }}^{\left( r \right)}}$, ${{\bf{p}}^{\left( r \right)}}$ and ${{\bm{\theta }}^{\left( r \right)}}$. Hence, we have 
\begin{equation}
	\begin{aligned}
		\Re \left( {{{\bf{Q}}^{\left( r \right)}},{{\bm{\psi }}^{\left( r \right)}},}\right.&\left.{{{\bm{\rho }}^{\left( r \right)}},{{\bf{p}}^{\left( r \right)}},{{\bm{\theta }}^{\left( r \right)}}} \right) \le\\ &\Re \left( {{{\bf{Q}}^{\left( {r + 1} \right)}},{{\bm{\psi }}^{\left( {r + 1} \right)}},{{\bm{\rho }}^{\left( r \right)}},{{\bf{p}}^{\left( r \right)}},{{\bm{\theta }}^{\left( r \right)}}} \right).
	\end{aligned}
\end{equation}
Similarly, in the step 4 of $\textbf{Algorithm 1}$, we can obtain the user PS ratio when ${{\bf{Q}}^{\left( r + 1\right)}}$, ${{\bm{\psi }}^{\left( r +1\right)}}$, ${{\bf{p}}^{\left( r \right)}}$ and ${{\bm{\theta }}^{\left( r \right)}}$ are given. Herein, we also have 
\begin{equation}
	\begin{aligned}
		\Re \left( {{{\bf{Q}}^{\left( {r + 1} \right)}},{{\bm{\psi }}^{\left( {r + 1} \right)}},}\right.&\left.{{{\bm{\rho }}^{\left( r \right)}},{{\bf{p}}^{\left( r \right)}},{{\bm{\theta }}^{\left( r \right)}}} \right) \le\\& \Re \left( {{{\bf{Q}}^{\left( {r + 1} \right)}},{{\bm{\psi }}^{\left( {r + 1} \right)}},{{\bm{\rho }}^{\left( {r + 1} \right)}},{{\bf{p}}^{\left( r \right)}},{{\bm{\theta }}^{\left( r \right)}}} \right).
	\end{aligned}
\end{equation}
In the step 5 of $\textbf{Algorithm 1}$, UAV transmit power allocation can be obtained when ${{\bf{Q}}^{\left( r + 1 \right)}}$, ${{{\bf{\psi }}^{\left( {r + 1} \right)}}}$, ${{\bm{\rho }}^{\left( r + 1\right)}}$ and ${{\bm{\theta }}^{\left( r \right)}}$ are given. Therefore, we have
\begin{equation}
	\begin{aligned}
		\Re \left( {{{\bf{Q}}^{\left( {r + 1} \right)}},{{\bm{\psi }}^{\left( {r + 1} \right)}},}\right.&\left.{{{\bm{\rho }}^{\left( {r + 1} \right)}},{{\bf{p}}^{\left( r \right)}},{{\bm{\theta }}^{\left( r \right)}}} \right) \le\\& \Re \left( {{{\bf{Q}}^{\left( {r + 1} \right)}},{{\bm{\psi }}^{\left( {r + 1} \right)}},{{\bm{\rho }}^{\left( {r + 1} \right)}},{{\bf{p}}^{\left( {r + 1} \right)}},{{\bm{\theta }}^{\left( r \right)}}} \right).
	\end{aligned}
\end{equation}
Finally, in the step 6 of $\textbf{Algorithm 1}$, IRS reflection coefficient can be obtained when ${{\bf{Q}}^{\left( r + 1 \right)}}$, ${{{\bf{\psi }}^{\left( {r + 1} \right)}}}$, ${{\bm{\rho }}^{\left( r + 1\right)}}$ and ${{\bf{p}}^{\left( r + 1\right)}}$ are fixed. Therefore, we have
\begin{equation}
	\begin{aligned}
		\Re \left( {{{\bf{Q}}^{\left( {r + 1} \right)}},}\right.&\left.{{{\bm{\psi }}^{\left( {r + 1} \right)}},{{\bm{\rho }}^{\left( {r + 1} \right)}},{{\bf{p}}^{\left( {r + 1} \right)}},{{\bm{\theta }}^{\left( r \right)}}} \right) \le\\& \Re \left( {{{\bf{Q}}^{\left( {r + 1} \right)}},{{\bm{\psi }}^{\left( {r + 1} \right)}},{{\bm{\rho }}^{\left( {r + 1} \right)}},{{\bf{p}}^{\left( {r + 1} \right)}},{{\bm{\theta }}^{\left( {r + 1} \right)}}} \right).
	\end{aligned}
\end{equation}
Based on the above, we can obtain
\begin{equation}
	\begin{aligned}
		\Re \left( {{{\bf{Q}}^{\left( r \right)}},}\right.&\left.{{{\bm{\psi }}^{\left( r \right)}},{{\bm{\rho }}^{\left( r \right)}},{{\bf{p}}^{\left( r \right)}},{{\bm{\theta }}^{\left( r \right)}}} \right) \le\\ &\Re \left( {{{\bf{Q}}^{\left( {r + 1} \right)}},{{\bm{\psi }}^{\left( {r + 1} \right)}},{{\bm{\rho }}^{\left( r+1 \right)}},{{\bf{p}}^{\left( r+1 \right)}},{{\bm{\theta }}^{\left( r +1\right)}}} \right).
	\end{aligned}
\end{equation}
which shows that the value of the objective function is non-decreasing after each iteration of $\textbf{Algorithm 1}$. Since the objective function is upper bounded by a finite value due to the limited transmit power of UAV, the convergence of $\textbf{Algorithm 1}$ can be guaranteed.

\section{Numerical results}
In this section, we verify the effectiveness of the proposed algorithm through the numerical results. In this paper, we consider that $K$ = 6 ground users are randomly distributed in a 500 $\times$ 500 circular area. We assume that the UAV flies at a fixed height $h_{u}$ = 100m, and its maximum flight speed $V_{max}$ = 20m/s. The maximum transmit power of UAV is $P_{max}$ = 43dBm. The coordinates of the initial horizontal position and final horizontal position of the UAV are (0, 250) and (500, 250), respectively. In addition, we consider that the IRS is fixed on a building with a height of $h_r$ = 30m, and its horizontal position coordinate is (250, 0). The number of IRS reflecting elements is $M$ = 20. The channel power gain at when the reference distance $d_0$ = 1m is $\beta_0$ = -30dB. We assume that the parameters of all users are the same, i.e., ${\xi _k} = 24{\rm{mW}}$, ${a_k} = 150$ and ${b_k} = 0.024$ \cite{8478252}. The Rice factor is $ \kappa_1 = \kappa_2$= = 3dB. The additive white Gaussian noise of the transmission channel is $\sigma$ = -80dBm. The path loss exponents of the UU-channel and IU-channel are $\alpha = \gamma = 2.2$. In addition, we set $\epsilon_max = 0.1$ and $\delta_{\max } = 5m$. The threshold of proposed algorithm is set as $10^{-3}$. In addition, the initialization method of the optimization variables is as follows: For the initialization of the user PS ratio, we make it randomly generated within $\left[ {0,1} \right]$. The initial value of the IRS reflection coefficient is randomly generated within $\left[ {0,2\pi } \right)$. The UAV transmit power is initialized by using an equal allocation scheme. The initial value of the UAV's trajectory is the position where the UAV is randomly generated in the coordinate plane in each time slot $n$, and the SIC decoding order is generated depending on the distance between the user and the UAV.

We first evaluate the convergence of the proposed algorithm. Fig. 3 shows the change of sum-rate with the number of iteration under different time periods $T$ and different IRS reflection elements $M$. We can see that the sum-rate increases rapidly with the number of iteration and can reach stable convergence in about six iterations, which verifies the convergence of the proposed algorithm. It can also be seen that when the IRS reflection elements are the same, a larger time duration $T$ can bring a larger system performance gain. Similarly, when the time duration $T$ is the same, more IRS reflection elements can also increase the system sum-rate.
\begin{figure}
	\centerline{\includegraphics[width=8cm]{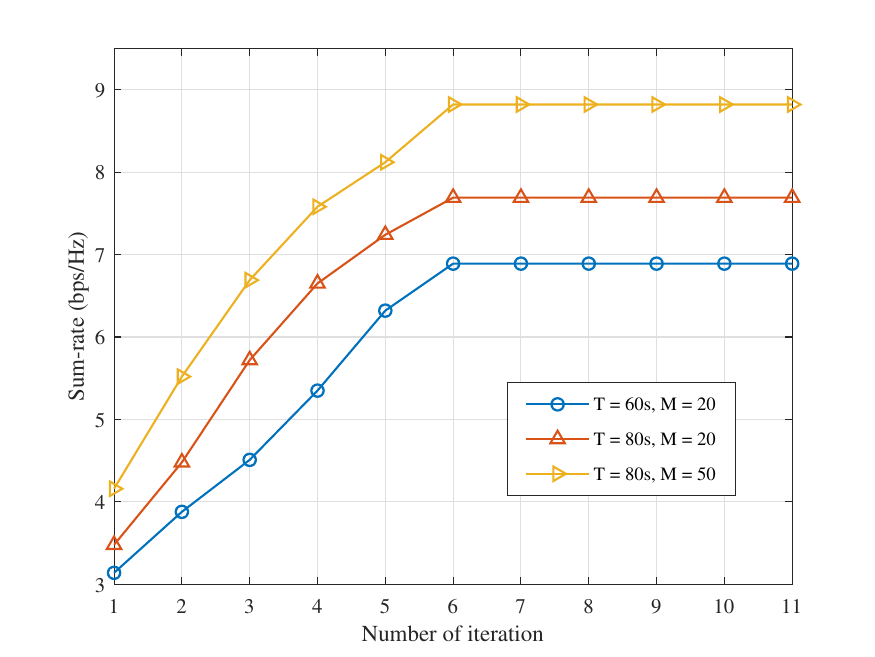}}
	\caption{Convergence behaviour of the proposed optimization algorithm.}
	\label{Fig3}
\end{figure}

Next, we elaborate the optimized trajectory of UAV at different time periods $T$ assisted by IRS in Fig. 4. From Fig. 4, we can see that when the time period $T$ is larger, UAV can be closer to more ground users, and can provide better wireless information transmission and wireless energy transmission. With the assistance of IRS, UAV can fly to IRS to balance the channel conditions of UU-channel and combined channels to provide better quality-of-service (QoS) for ground users. In addition, when the time period $T$ is large enough, the UAV can spend more time staying near the IRS to provide higher quality services for ground users.
\begin{figure}
	\centerline{\includegraphics[width=8cm]{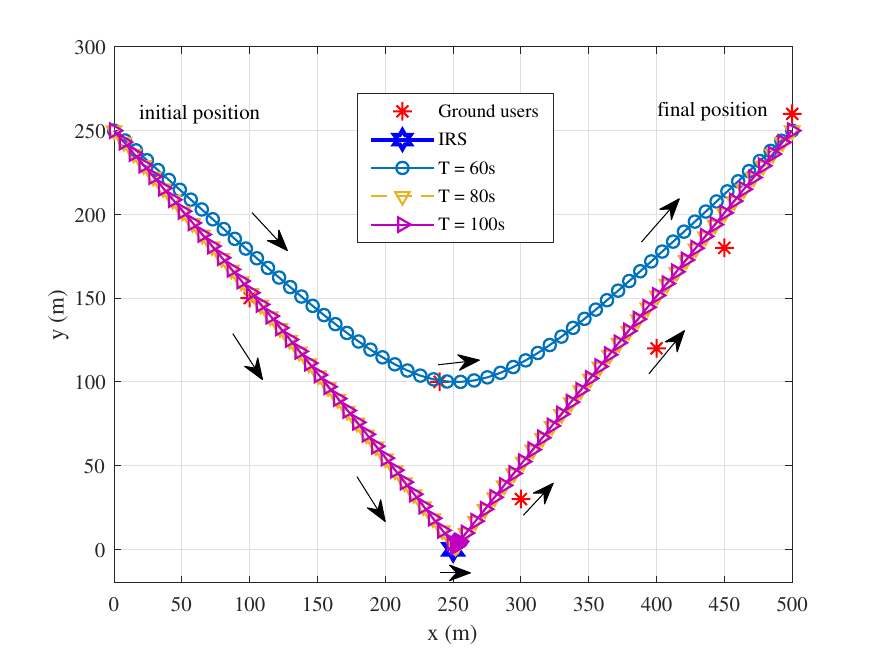}}
	\caption{UAV optimization trajectory at different time periods $T$ assisted by IRS.}
	\label{Fig4}
\end{figure}

\begin{figure}
	\centerline{\includegraphics[width=8cm]{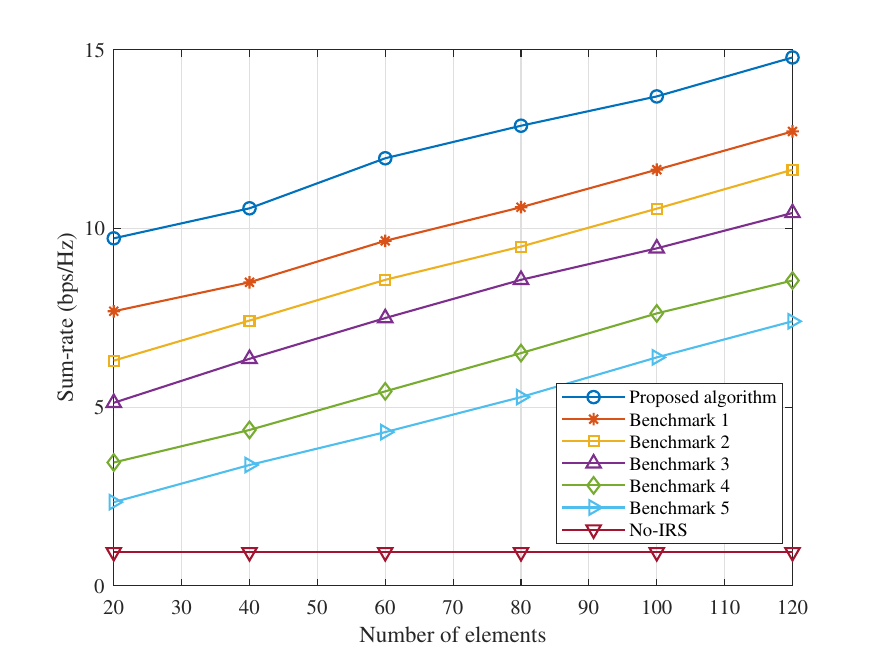}}
	\caption{Sum-rate versus the number of IRS reflection elements for the proposed algorithm and different benchmark algorithms.}
	\label{Fig5}
\end{figure}	
Then, we compare the proposed algorithm with several other benchmark algorithms as follows: (1) Benchmark 1 (Equ-power): UAV transmit power allocation adopts the scheme of equal allocation. The optimization scheme for other variables is the same as Algorithm 1. (2) Benchmark 2 (No-opt-trajectory): The UAV trajectory is not optimized, and a random scheme is used. The optimization scheme for other variables is the same as Algorithm 1. (3) Benchmark 3 (No-opt-phase): The IRS phase shift is not optimized and a random phase shift is used. The optimization scheme for other variables is the same as Algorithm 1. (4) Benchmark 4 (Com-opt): All optimization variables are optimized only once using the optimization algorithm for each sub-problem in Algorithm 1, and no alternate optimization is performed. (5) Benchmark 5 (Ran-opt): All optimization variables are random. (6) Benchmark 6 (Static-opt): Optimize the deployment of static UAV. (7) Benchmark 7 (Static-ran): Random deployment of static UAV.  For benchmark 6 and benchmark 7, the optimization for other variables is the same as the proposed algorithm 1 except that UAV trajectory optimization is not considered. They consider the static deployment problem of UAV. Benchmark 6 considers the UAV to be deployed at a random position in the circular area, and benchmark 7 uses the exhaustive method to find the sub-optimal deployment position of the circular area after optimizing other variables.  (8) Benchmark 8 (Non-max-rate): UAV without maximum flight rate constraint. (9) Benchmark 9 (Str-trajectory): UAV flies in a straight line from the initial position to the final position. (10) No-IRS: Without the assistance of IRS, the optimization algorithm for other variables is the same as Algorithm 1.

Fig. 5 shows the variation of the system sum-rate with the number of IRS reflection elements for the proposed algorithm and benchmark algorithms. It can be seen that when the number of IRS reflection elements increases, the performance of the proposed algorithm improves. This is because the number of IRS reflection elements increases, the number of combined channels will also increase, which can provide better channel quality for ground users, i.e., the sum-rate will also increase accordingly. Compared with other benchmark algorithms, our proposed algorithm has obvious performance gains. Specifically, when the number of IRS reflection elements is the same, the performance of the proposed algorithm is better than that of benchmark 1, which is mainly due to the fact that in benchmark 1, the UAV transmit power is not optimized, but an equal allocation scheme is adopted. Similarly, the main reason why the proposed algorithm outperforms benchmark 2 and benchmark 3 is that the latter two are not optimized for the UAV trajectory and the phase shift of IRS, respectively. In addition, the performance of benchmark 2 is better than that of benchmark 3, indicating that the gain of the proposed algorithm mainly comes from the phase shift optimization of IRS. This is because if the phase shift setting of the IRS is unreasonable, it is likely to deteriorate the system performance. Moreover, the performance of the proposed algorithm is better than benchmark 4, mainly because benchmark 4 does not have alternate optimization to achieve global convergence. Benchmark 5 has the worst performance due to its random scheme. Finally, it can be seen that in terms of system performance, the IRS-assisted system has a larger gain than the non-IRS-assisted system, because the IRS can improve the system performance by increasing the directional beam. Therefore, it is practical to improve the performance of conventional UAV SWIPT network by this low-cost passive IRS.

In Fig. 6, we compare the change of sum-rate with the UAV height $h_u$ for the proposed algorithm and benchmark algorithms. It describes that when the height of the UAV increases, the system sum-rate decreases. This is mainly due to when the UAV is close to the ground users, the quality of the air-ground channel provided can be improved, thus the user's rate can also be increased, thereby enhancing the system sum-rate. When the height of the UAV is the same, the performance of our proposed algorithm still has a significant performance gain. The reasons for generating the gain are similar to those mentioned above, and will not be repeated here.

\begin{figure}
	\centerline{\includegraphics[width=8cm]{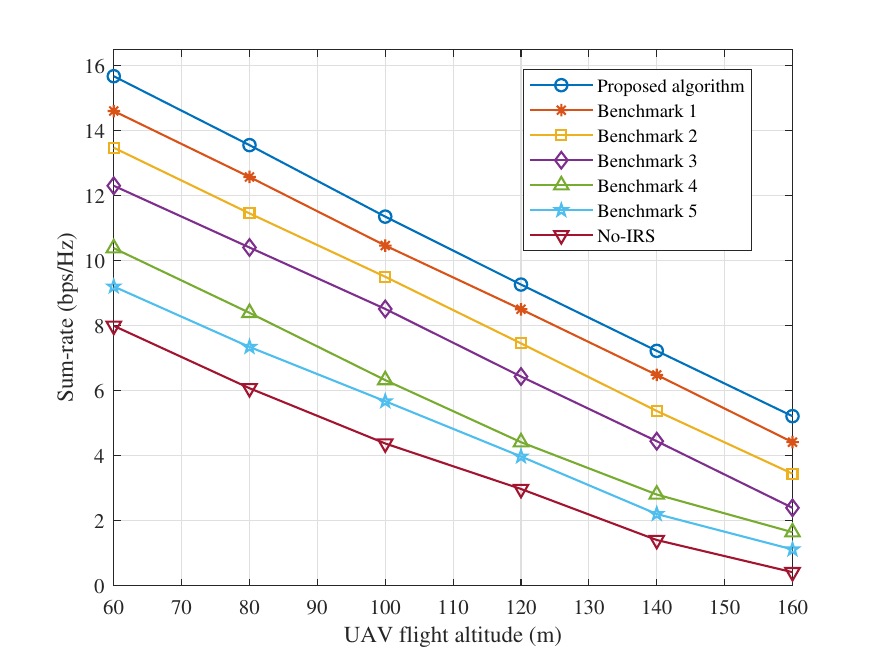}}
	\caption{Sum-rate versus the UAV flight altitude for the proposed algorithm and different benchmark algorithms.}
	\label{Fig6}
\end{figure}

Fig. 7 shows the effect of UAV on system sum-rate with dynamic flight and static deployment. It can be seen that compared with benchmark 7, benchmark 6 has better performance in terms of sum-rate. Moreover, the proposed algorithm considers the UAV's optimized trajectory, i.e., the UAV can fly dynamically and serve as many ground users as possible. Therefore, the performance of the proposd algorithm has a significant gain compared to benchmark 6 and benchmark 7. In addition, benchmark 3 considers the equal allocation of UAV transmit power, i.e., for all users to distribute the same power, the rate of users farther from UAV will be reduced. Therefore, the proposed algorithm considering UAV power allocation optimization has a gain in terms of sum-rate compared to benchmark 3. Next, we consider the case that the UAV without maximum flight rate constraint, i.e., benchmark 8. In benchmark 8, the UAV can hover directly above each ground user long enough without considering the flight rate constraints, and then fly to the next user, which can significantly improve the throughput of ground users. Therefore, compared with the proposed algorithm, the performance of this algorithm in terms of sum-rate is better. However, in practical scenarios, UAV usually has a maximum flight rate constraint. In summary, the proposed algorithm is closer to the practical setup and has a higher performance gain compared to several benchmark algorithms. 
\begin{figure}
	\centerline{\includegraphics[width=8cm]{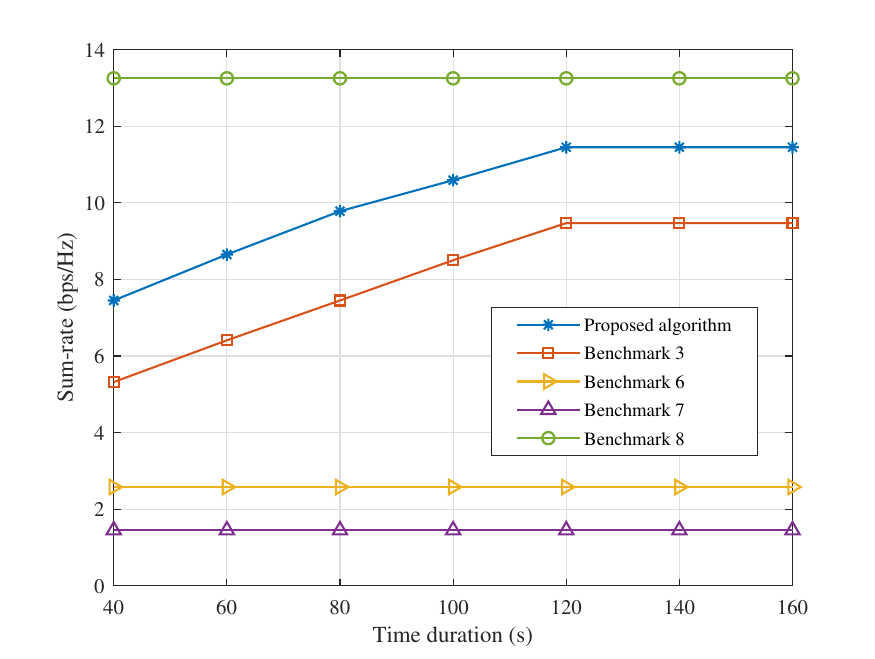}}
	\caption{Sum-rate versus time period $T$ for the proposed algorithm and benchmark algorithms.}
	\label{Fig7}
\end{figure}

Next, Fig. 8 illustrates the effect of different UAV trajectory schemes in terms of sum-rate. benchmark 8 has been mentioned above, here we can understand it as a UAV trajectory scheme, i.e., UAV provides services to ground users through hover-flight. UAV flies directly above each user to provide services, and then flies to another user. Therefore, compared to the proposed algorithm, the rate of each user can be improved, thereby the system sum-rate can be improved. In benchmark 9, UAV does not fly according to the user's position, and not all users' rates are guaranteed well, so the system sum-rate will decrease compared to the proposed algorithm.
\begin{figure}
	\centerline{\includegraphics[width=8cm]{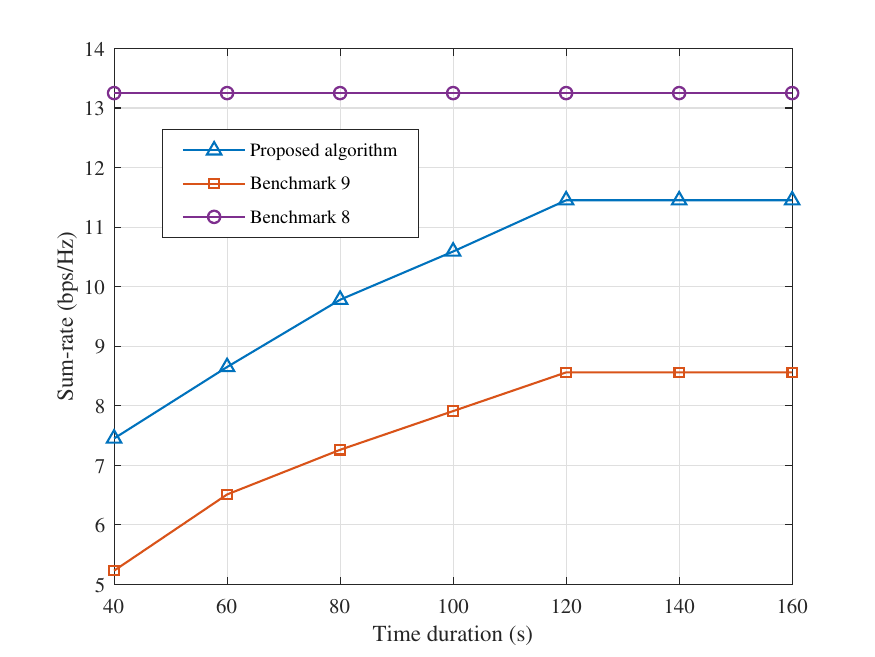}}
	\caption{Sum-rate versus time period $T$ for the proposed algorithm and different UAV trajectory designs.}
	\label{Fig8}
\end{figure}

In Fig. 9, the sum-rate versus UAV maximum transmit power for the proposed algorithm and benchmark algorithms is shown. It can be seen that when the UAV maximum transmit power increases, the sum-rate will also increase, which can be explained as the increase of the UAV transmit power, the rate of all users can be improved, so sum-rate of system will also increase. In addition, the performance of our proposed algorithm is superior to the benchmark algorithms, mainly because we consider the joint optimization of all variables and achieve the convergence of the problem by applying AO technique.
\begin{figure}
	\centerline{\includegraphics[width=8cm]{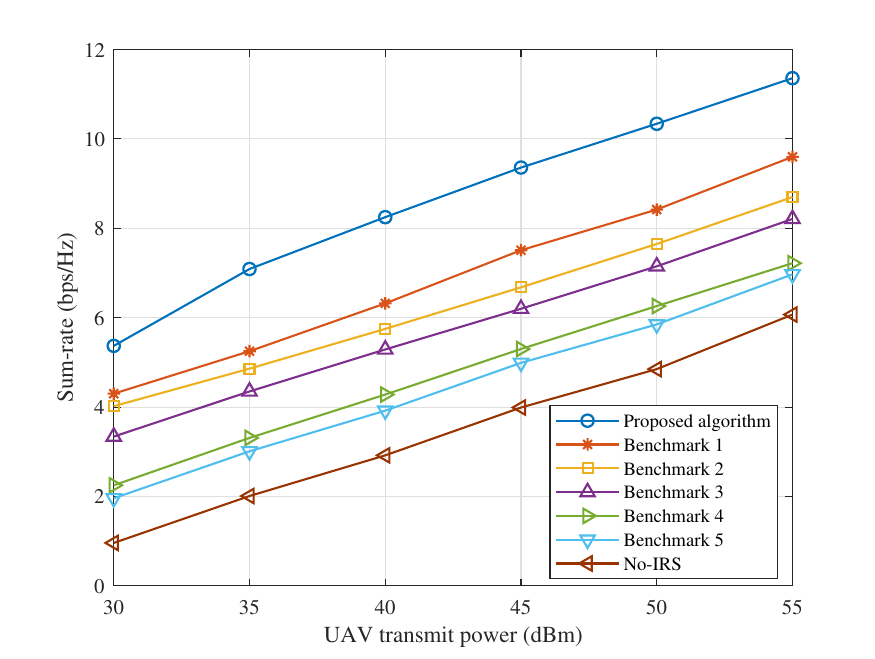}}
	\caption{Sum-rate versus UAV maximum transmit power for the proposed algorithm and benchmark algorithms.}
	\label{Fig9}
\end{figure}

Finally, in Fig. 10, we depict the change of sum-rate with the user energy harvesting threshold under different UAV maximum flight speed. It can be seen from Fig. 10 that when the user energy harvesting threshold is fixed, the maximum flight speed of UAV has an impact on the system sum-rate, i.e., the sum-rate increases as the UAV maximum flight rate increases. This is because the greater the  UAV maximum flight speed, the longer UAV can hover at the user, so it can provide better QoS for the users. In addition, considering the SWIPT architecture in this paper, the change of sum-rate with the energy harvesting threshold of ground users reflects the trade-off in the IRS-empowered UAV SWIPT networks. When the user energy harvesting threshold continues to increase, i.e., the user's energy requirement becomes greater, the sum-rate of system will gradually decrease. The main reason is that the increase in energy requirements of ground users triggers the power splitting ratio to use more power resources of UAV for wireless energy transmission and a smaller portion for wireless information transmission, which leads to a reduction in the rate of ground users, thereby the system sum-rate will decrease.
\begin{figure}
	\centerline{\includegraphics[width=8cm]{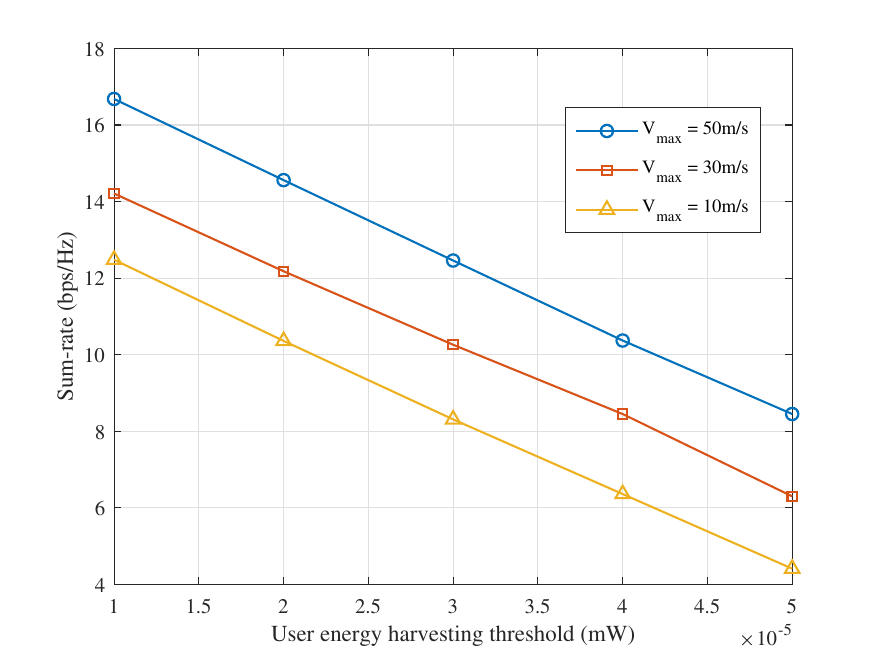}}
	\caption{Sum-rate versus user energy harvesting threshold with different UAV maximum flight speed.}
	\label{Fig9}
\end{figure}
\section{Conclusion}
This paper investigates the sum-rate maximization problem of IRS empowered UAV SWIPT networks. Specifically, under the constraints of the energy harvesting threshold, UAV trajectory, SIC decoding order, UAV transmit power allocation, PS ratio and IRS reflection coefficient are jointly optimized. First, we transform the problem into a tractable problem. Then, in order to solve the transformed problem, we apply the AO algorithm framework to divide the original problem into four sub-problems for solving. Specifically, when the other three sets of variables are given, we apply SCA, penalty function method and DC programming to alternately optimize the optimization variables until convergence is achieved. Then, the computational complexity and convergence analysis of the proposed algorithm is given. Finally, the numerical simulation results verify the convergence and effectiveness of the algorithm, which shows that the proposed algorithm can significantly improve the sum-rate of the system, and the role of IRS is extremely important, and the system performance can be improved at a lower cost, which is very meaningful.

\ifCLASSOPTIONcaptionsoff
  \newpage
\fi


\bibliographystyle{IEEEtran}
\bibliography{reference}

\end{document}